\documentclass[a4paper,12pt]{article}
\usepackage{amssymb}

\usepackage{amsmath}
\usepackage{bm}
\usepackage{boxedminipage}
\usepackage[pdftex]{graphicx}
\usepackage{graphicx}
\usepackage{enumerate}

\makeatletter \@addtoreset{equation}{section}

\makeatletter\renewcommand\section{\@startsection {section}{1}{\z@}
{-3.5ex \@plus -1ex \@minus -.2ex}
{2.3ex \@plus.2ex}
{\normalfont\large\bfseries}}
\renewcommand\subsection{\@startsection{subsection}{2}{\z@}                                     {-3.25ex\@plus -1ex \@minus -.2ex}                                     {1.5ex \@plus .2ex}                                     {\normalfont\bfseries}}
\def\Label#1{\label{#1}  \smash{\hbox to0pt{\raise1ex\hbox{\tiny[#1]}\hss}}}
\def\noLabels{\let\Label=\label}
\def\nobbibitem{\let\bbibitem=\bibitem}

\begin{document}

\begin{titlepage}

    \thispagestyle{empty}
    \begin{flushright}
        \hfill{UCB-PTH-09/05}\\
        \hfill{LBNL-1936E}\\
        \hfill{CERN-PH-TH/2009-001}\\
        \hfill{SU-ITP-09/04}\\
    \end{flushright}

    \begin{center}
        { \Huge{\textbf{Duality, Entropy and ADM Mass\\ \vspace{5pt}in Supergravity}}}

        \vspace{10pt}

        {\large{{\bf Bianca L. Cerchiai$^{\spadesuit,\diamondsuit}$, \ Sergio Ferrara$^{\clubsuit,\flat}$,\\\ Alessio Marrani$^{\heartsuit,\flat}$ and \ Bruno Zumino$^{\spadesuit,\diamondsuit}$}}}

        \vspace{20pt}

        {$\spadesuit$ \it Lawrence Berkeley National Laboratory, \\
        Theory Group, Bldg 50A5104\\
        1 Cyclotron Rd, Berkeley, CA 94720-8162, USA\\
        \texttt{BLCerchiai@lbl.gov, zumino@thsrv.lbl.gov}}

        \vspace{5pt}

        {$\diamondsuit$ \it Department of Physics, University of California,\\
        Berkeley, CA 94720-8162, USA}

        \vspace{5pt}

        {$\clubsuit$ \it Theory division, CERN, Geneva, Switzerland \\
        CH 1211, Geneva 23, Switzerland\\
        \texttt{sergio.ferrara@cern.ch}}

        \vspace{5pt}

         {$\flat$ \it INFN - LNF, \\
          Via Enrico Fermi 40, I-00044 Frascati, Italy\\
         \texttt{marrani@lnf.infn.it}}



        \vspace{5pt}

        {$\heartsuit$ \it Stanford Institute for Theoretical Physics\\
        Department of Physics, 382 Via Pueblo Mall, Varian Lab,\\
        Stanford University, Stanford, CA 94305-4060, USA}


        \vspace{30pt}

        {ABSTRACT}

    \end{center}

    \vspace{5pt}

 We consider the Bekenstein-Hawking entropy-area formula in four dimensional
 extended ungauged supergravity and its electric-magnetic duality property.

Symmetries of both ``large" and ``small" extremal black holes are
considered, as well as the ADM mass formula for $\mathcal{N}=4$ and
$\mathcal{N}=8$ supergravity, preserving different fraction of
supersymmetry.

The interplay between BPS conditions and duality properties is an
important aspect of this investigation.


\end{titlepage}
\newpage\tableofcontents

\section{\label{Intro}Introduction}

In $d=4$ extended \textit{ungauged} supergravity theories based on
scalar manifolds which are (\textit{at least locally})
\textit{symmetric} spaces
\begin{equation}
M=\frac{G}{H},  \label{G/H}
\end{equation}
it is known that the classification of static, spherically symmetric and
asymptotically flat extremal black hole (BH) solutions is made in terms of
\textit{charge orbits} of the corresponding \textit{classical}
electric-magnetic duality group group $G$ \cite
{GZ,F-Scherk-Z-1,FM,FG1,LPS,BFGM1} (later called $U$-duality\footnote{%
Here $U$-duality is referred to as the \textit{``continuous''} version,
valid for large values of the charges, of the $U$-duality groups introduced
by Hull and Townsend \cite{HT}.} in string theory) .

These orbits correspond to certain values taken by a \textit{duality
invariant}\footnote{%
By \textit{duality invariant}, throughout our treatment we mean that such a
combination is $G$-invariant. Thus, it is actually \textit{independent} on
the scalar fields, and it depends only on \textit{``bare''} electric and
magnetic (asymptotical) charges (defined in Eq. (\ref{P})).} combination of
the \textit{``dressed''} central charges and matter charges. Denoting such
an invariant by $\mathcal{I}$, the set of scalars parametrizing the
symmetric manifold $M$ by $\phi $, and the set of \textit{``bare''} magnetic
and electric charges of the (\textit{dyonic}) BH configuration by the $%
2n\times 1$ symplectic vector
\begin{equation}
\mathcal{P}\equiv \left(
\begin{array}{c}
p^{\Lambda } \\
\\
q_{\Lambda }
\end{array}
\right) ,~\Lambda =1,...,n,  \label{P}
\end{equation}
then it holds that
\begin{equation}
\partial _{\phi }\mathcal{I}\left( \phi ,\mathcal{P}\right)
=0\Leftrightarrow \mathcal{I}=\mathcal{I}\left( \mathcal{P}\right) .
\label{U-inv}
\end{equation}
In some cases, the relevant invariant $\mathcal{I}$ is not enough to
characterize the orbit, and additional constraints are needed. This is
especially the case for the so-called\footnote{%
Throughout the present treatment, we will respectively call \textit{small}
or \textit{large} (extremal) BHs those BHs with vanishing or
non-vanishing area of the event horizon (and therefore with vanishing or
non-vanishing \textit{Bekenstein-Hawking entropy} \cite{Bek-Hawking}). For
symmetric geometries, they can be $G$-invariantly characterized respectively
by $\mathcal{I}=0$ or by $\mathcal{I}\neq 0$.} \textit{``small''} BHs, in
which case $\mathcal{I}=0$ on the corresponding orbit \cite
{FM,FG1,DFL-0-brane}.

An explicit expression for the $E_{7\left( 7\right) }$-invariant \cite
{Cartan} was firstly introduced in supergravity in \cite{CJ}, and then
adopted in the study of BH entropy in \cite{KK}. The additional $U$%
-invariant constraints which specify charge orbits with higher supersymmetry
were given in \cite{FM}. The corresponding (\textit{large} and \textit{%
small}) charge orbits for $\mathcal{N}=8$ and exceptional $\mathcal{N}=2$
supergravity were determined in \cite{FG1}, whereas the \textit{large}
orbits for all other \textit{symmetric} $\mathcal{N}=2$ supergravities were
obtained in \cite{BFGM1}, and then in \cite{ADFT-review} for all $\mathcal{N}%
>2$-extended theories. Furthermore, the invariant for $\mathcal{N}=4$
supergravity was earlier discussed in \cite{CY,DLR}.

The invariants play an important role in the \textit{attractor mechanism}
\cite{FKS,Strominger-1,FK1,FK2,FGK}, because the Bekenstein-Hawking BH
entropy \cite{Bek-Hawking}, determined by evaluating the \textit{effective
black hole potential} (\cite{FK1,FK2,FGK})
\begin{equation}
V_{BH}\left( \phi ,\mathcal{P}\right) \equiv -\frac{1}{2}\mathcal{P}^{T}%
\mathcal{M}\left( \phi \right) \mathcal{P}  \label{VBH-1}
\end{equation}
at its critical points, actually coincides with the relevant invariant:
\begin{equation}
\frac{S_{BH}}{\pi }=\left. V_{BH}\right| _{\partial _{\phi
}V_{BH}=0}=V_{BH}\left( \phi _{H}\left( \mathcal{P}\right) ,\mathcal{P}%
\right) =\left| \mathcal{I}\left( \mathcal{P}\right) \right| ^{1/2}~\text{(%
\textit{or~}}\left| \mathcal{I}\left( \mathcal{P}\right) \right| \text{)}.
\label{SBH-1}
\end{equation}
In Eq. (\ref{VBH-1}) $\mathcal{M}$ stands for the $2n\times 2n$ real
(negative definite) symmetric scalar-dependent symplectic matrix
\begin{equation}
\mathcal{M}\left( \phi \right) \equiv \left(
\begin{array}{ccc}
Im\mathcal{N}_{\Lambda \Sigma }+Re\mathcal{N}_{\Lambda \Xi }\left( Im%
\mathcal{N}\right) ^{-1\mid \Xi \Delta }Re\mathcal{N}_{\Delta \Sigma } & ~~
& -Re\mathcal{N}_{\Lambda \Xi }\left( Im\mathcal{N}\right) ^{-1\mid \Xi
\Sigma } \\
~~ & ~~ & ~~ \\
-\left( Im\mathcal{N}\right) ^{-1\mid \Lambda \Delta }Re\mathcal{N}_{\Xi
\Sigma } & ~~~~ & \left( Im\mathcal{N}\right) ^{-1\mid \Lambda \Sigma }
\end{array}
\right) ,  \label{M}
\end{equation}
defined in terms of the normalization of the Maxwell and topological terms%
\footnote{%
Attention should be paid in order to distinguish between the notations of
the number $\mathcal{N}$ of \textit{supercharges} of a supergravity theory
and the \textit{kinetic vector matrix} $\mathcal{N}_{\Lambda \Sigma }$
introduced in Eqs. (\ref{M}) and (\ref{N}).}
\begin{equation}
Im\mathcal{N}_{\Lambda \Sigma }\left( \phi \right) F^{\Lambda }F^{\Sigma
},~~~Re\mathcal{N}_{\Lambda \Sigma }\left( \phi \right) F^{\Lambda }\tilde{F}%
^{\Sigma }  \label{N}
\end{equation}
of the corresponding supergravity theory (see \textit{e.g.} \cite
{ADF-central,CDF-review} and Refs. therein). Furthermore, in Eq. (\ref{SBH-1}%
) $\phi _{H}\left( \mathcal{P}\right) $ denotes the set of charge-dependent,
stabilized horizon values of the scalars, solutions of the criticality
conditions for $V_{BH}$:
\begin{equation}
\left. \frac{\partial V_{BH}\left( \phi ,\mathcal{P}\right) }{\partial \phi }%
\right| _{\phi =\phi _{H}\left( \mathcal{P}\right) }\equiv 0.
\end{equation}

For the case of charge orbits corresponding to \textit{small} BHs, in
the case of a \textit{single-center} solution $\mathcal{I}\left( \mathcal{P}%
\right) =0$, and thus the event horizon area vanishes, and the solution is
singular (\textit{i.e.} with vanishing Bekenstein-Hawking entropy). However,
the charge orbits with vanishing duality invariant play a role for \textit{%
multi-center} solutions as well as for elementary BH constituents through
which \textit{large} (\textit{i.e.} with non-vanishing
Bekenstein-Hawking entropy) BHs are made \cite{GLS,BGL,DGSVY}.\smallskip

In the present investigation, we re-examine the duality invariant and the $U$%
-invariant classification of charge orbits of $\mathcal{N}=8$, $d=4$
supergravity, we give a complete analysis of the $\mathcal{N}=4$ \textit{%
large }and \textit{small }charge orbits, and we also derive a
diffeomorphism-invariant expression of the $\mathcal{N}=2$ duality
invariant, which is common to all symmetric spaces and which is completely
independent on the choice of a symplectic basis.\medskip

The paper is organized as follows.

In Sect. \ref{Duality} we recall some basic facts about electric-magnetic
duality in $\mathcal{N}$-extended supergravity theories, firstly treated in
\cite{F-Scherk-Z-1}. The treatment follows from the general analysis of \cite
{GZ}, and the dictionary between that paper and the present work is given.

In Sect. \ref{N=8-ungauged} we re-examine $\mathcal{N}=8$, $d=4$
supergravity and the $E_{7\left( 7\right) }$-invariant
characterization of its charge orbits. This refines, re-organizes
and extends the various results of \cite{FM,FG1,LPS,DFL-0-brane}.

In Sect. \ref{N=4-ungauged} we reconsider \textit{matter coupled} $\mathcal{N%
}=4$, $d=4$ supergravity. The $SL\left( 2,%
\mathbb{R}\right) \times SO\left( 6,M\right) $-invariant
characterization of all its BPS and non-BPS charge orbits, firstly
obtained in \cite{FM,DFL-0-brane}, is the starting point of the
novel results presented in this Section.

Sect. \ref{N=2-symmetric-ungauged} is devoted to the analysis of the $%
\mathcal{N}=2$, $d=4$ case \cite{FM}. Beside the generalities on the
special K\"{a}hler geometry of Abelian vector multiplets' scalar
manifold, the results of this Section are novel. In particular, a
formula for the duality invariant is determined, which is
\textit{diffeomorphism-invariant} and holds true for all
\textit{symmetric} special K\"{a}hler manifolds (see \textit{e.g.}
\cite {dWVVP} and Refs. therein), regardless of the considered
symplectic basis.

Sect. \ref{ADM-Mass}, starting from the analysis of \cite
{FM,DFL-0-brane}, deals with the issue of the \textit{ADM mass}
\cite{ADM} in $\mathcal{N}=8$ (Subsect. \ref{N=8-ADM-Mass}) and
$\mathcal{N}=4$ (Subsect. \ref{N=4-ADM-Mass}), ungauged $d=4$
supergravities. In general, for all supersymmetric orbits the
\textit{ADM mass} has a known explicit expression, depending on the
number of supersymmetries preserved by the state which is supported
by the considered orbit (saturating the \textit{BPS}
\cite{BPS}\textbf{\ }bound).

\section{\label{Duality}Electric-Magnetic Duality in Supergravity : Basic
Facts}

The basic requirement for consistent coupling of a non-linear sigma model
based on a \textit{symmetric} manifold (\ref{G/H}) to $\mathcal{N}$%
-extended, $d=4$ supergravity (see \textit{e.g.} \cite{ADF-central} and
Refs. therein) is that the vector field strengths and their duals (through
\textit{Legendre transform} with respect the Lagrangian density $\mathcal{L}$%
)
\begin{equation}
F^{\Lambda },~~G_{\Lambda }\equiv \frac{\delta \mathcal{L}}{\delta
F^{\Lambda }},
\end{equation}
belong to a \textit{symplectic} representation $\mathbf{R}_{s}$ of the
global (\textit{classical}, see Footnote 1) $U$-duality group $G$, given by $%
2n\times 2n$ matrices with block structure
\begin{equation}
\left(
\begin{array}{ccc}
A &  & B \\
&  &  \\
C &  & D
\end{array}
\right) \in Sp\left( 2n,\mathbb{R}\right) ,  \label{def-1}
\end{equation}
where $A$, $B$, $C$ and $D$ are $n\times n$ real matrices. By defining the $%
2n\times 2n$ \textit{symplectic} metric (each block being $n\times n$)
\begin{equation}
\Omega \equiv \left(
\begin{array}{ccc}
0 &  & -1 \\
&  &  \\
1 &  & 0
\end{array}
\right) ,  \label{Omega}
\end{equation}
the \textit{finite symplecticity condition} for a $2n\times 2n$ real matrix $%
P$
\begin{equation}
P^{T}\Omega P=\Omega
\end{equation}
yields the following relations to hold for the block components of the
matrix defined in Eq. (\ref{def-1}):
\begin{eqnarray}
A^{T}C-C^{T}A &=&0;  \label{sympl-1} \\
B^{T}D-D^{T}B &=&0;  \label{sympl-2} \\
A^{T}D-C^{T}B &=&1.  \label{sympl-3}
\end{eqnarray}
An analogous, equivalent definition of the representation $\mathbf{R}_{s}$
is the following one: $\mathbf{R}_{s}$ is real and it contains the singlet
in its $2$-fold antisymmetric tensor product
\begin{equation}
\left( \mathbf{R}_{s}\times \mathbf{R}_{s}\right) _{a}\ni \mathbf{1}.
\label{eq-sympl}
\end{equation}

If the basic requirements (\ref{sympl-1})-(\ref{sympl-3}) or (\ref{eq-sympl}%
) are met, the \textit{coset representative} of $M$ in the symplectic
representation $\mathbf{R}_{s}$ is given by the (\textit{scalar-dependent}) $%
2n\times 2n$ matrix
\begin{equation}
S\left( \phi \right) \equiv \left(
\begin{array}{ccc}
A\left( \phi \right) &  & B\left( \phi \right) \\
&  &  \\
C\left( \phi \right) &  & D\left( \phi \right)
\end{array}
\right) \in Sp\left( 2n,\mathbb{R}\right) .  \label{CERNN-1}
\end{equation}
A particular role is played by the two (\textit{scalar-dependent}) complex $%
n\times n$ matrices $f$ and $h$, which do satisfy the properties
\begin{eqnarray}
-f^{\dag }h+h^{\dag }f &=&i1,  \label{f-h-1} \\
-f^{T}h+h^{T}f &=&0.  \label{f-h-2}
\end{eqnarray}
The constraining relations (\ref{f-h-1}) and (\ref{f-h-2}) are equivalent to
require that
\begin{equation}
S\left( \phi \right) =\sqrt{2}\left(
\begin{array}{ccc}
Ref &  & -Imf \\
&  &  \\
Reh &  & -Imh
\end{array}
\right) ,
\end{equation}
or equivalently:
\begin{eqnarray}
f &=&\frac{1}{\sqrt{2}}\left( A-iB\right) ;~ \\
h &=&\frac{1}{\sqrt{2}}\left( C-iD\right) .  \label{CERNN-2}
\end{eqnarray}

In order to make contact with the formalism introduced by Gaillard and
Zumino in \cite{GZ}, it is convenient to use another (complex) basis, namely
the one which maps an element $S\in Sp\left( 2n,\mathbb{R}\right) $ into an
element $U\in U\left( n,n\right) \cap $ $Sp\left( 2n,\mathbb{C}\right) $.
The change of basis is exploited through the matrix
\begin{equation}
\mathcal{A}\equiv \frac{1}{\sqrt{2}}\left(
\begin{array}{ccc}
1 &  & 1 \\
&  &  \\
-i1 &  & i1
\end{array}
\right) ,~~\mathcal{A}^{-1}=\mathcal{A}^{\dag }.
\end{equation}
The (\textit{scalar-dependent}) matrix $U$ is thus defined as follows:
\begin{equation}
U\left( \phi \right) \equiv \mathcal{A}^{-1}S\mathcal{A}=\frac{1}{\sqrt{2}}%
\left(
\begin{array}{ccc}
f+ih &  & \overline{f}+i\overline{h} \\
&  &  \\
f-ih &  & \overline{f}-i\overline{h}
\end{array}
\right) \in U\left( n,n\right) \cap Sp\left( 2n,\mathbb{C}\right) .
\end{equation}
This is the matrix named $S$ in Eq. (5.1) of \cite{GZ}. Correspondingly, the
$Sp\left( 2n,\mathbb{R}\right) $-covariant vector $\left( F^{\Lambda
},G_{\Lambda }\right) ^{T}$ is mapped into the vector
\begin{equation}
\mathcal{A}^{-1}\left(
\begin{array}{c}
F^{\Lambda } \\
\\
G_{\Lambda }
\end{array}
\right) =\frac{1}{\sqrt{2}}\left(
\begin{array}{ccc}
1 &  & i1 \\
&  &  \\
1 &  & -i1
\end{array}
\right) \left(
\begin{array}{c}
F^{\Lambda } \\
\\
G_{\Lambda }
\end{array}
\right) =\frac{1}{\sqrt{2}}\left(
\begin{array}{c}
F^{\Lambda }+iG_{\Lambda } \\
\\
F^{\Lambda }-iG_{\Lambda }
\end{array}
\right) .
\end{equation}
The kinetic vector matrix $\mathcal{N}_{\Lambda \Sigma }$ appearing in Eqs. (%
\ref{M}) and (\ref{N}) is given by (in matrix notation)
\begin{equation}
\mathcal{N}\left( \phi \right) =hf^{-1}=\left( f^{-1}\right) ^{T}h^{T},
\end{equation}
and it is named $-i\overline{K}$ in \cite{GZ}.

Thus, by introducing the $2n\times 1$ ($n\times n$ \textit{matrix-valued})
complex vector
\begin{equation}
\Xi \equiv \left(
\begin{array}{c}
f \\
\\
h
\end{array}
\right)
\end{equation}
and recalling the definition (\ref{M}), the matrix $\mathcal{M}$ can be
written as
\begin{eqnarray}
\mathcal{M}\left( \phi \right) &=&-i\Omega +2\Omega \Xi \left( \Omega \Xi
\right) ^{\dag }=-i\Omega -2\Omega \Xi \Xi ^{\dag }\Omega =  \notag \\
&=&-i\Omega -2\left(
\begin{array}{c}
-h \\
\\
f
\end{array}
\right) \left(
\begin{array}{ccc}
h^{\dag }, &  & -f^{\dag }
\end{array}
\right) =  \notag \\
&=&-i\left(
\begin{array}{ccc}
0 &  & -1 \\
&  &  \\
1 &  & 0
\end{array}
\right) +2\left(
\begin{array}{ccc}
hh^{\dag } &  & -hf^{\dag } \\
&  &  \\
-fh^{\dag } &  & ff^{\dag }
\end{array}
\right) .  \label{M-2}
\end{eqnarray}
Eqs. (\ref{VBH-1}), (\ref{M}) and (\ref{M-2}) imply that
\begin{eqnarray}
V_{BH}\left( \phi ,\mathcal{P}\right) &\equiv &-\frac{1}{2}\mathcal{P}^{T}%
\mathcal{M}\left( \phi \right) \mathcal{P}=Tr\left(\mathcal{ZZ}^{\dag
}\right)=Tr\left( \mathcal{Z}^{\dag }\mathcal{Z}\right) =  \notag \\
&=&\sum_{A>B=1}^{\mathcal{N}}Z_{AB}\overline{Z}^{AB}+Z_{I}\overline{Z}^{I}=%
\frac{1}{2}Z_{AB}\overline{Z}^{AB}+Z_{I}\overline{Z}^{I}=  \notag \\
&=&\frac{1}{2}Tr\left( ZZ^{\dag }\right) +Z_{I}\overline{Z}^{I}=\frac{1}{2}%
Tr\left( Z^{\dag }Z\right) +Z_{I}\overline{Z}^{I},  \label{VBH-3}
\end{eqnarray}
where ($A$, $B=1,...,\mathcal{N}$ and $I=1,...,m$ throughout; recall $%
\Lambda =1,...,n$)
\begin{gather}
\mathcal{Z}\equiv \mathcal{P}^{T}\Omega \Xi =qf-ph=\left( Z_{AB}\left( \phi ,%
\mathcal{P}\right) ,~Z_{I}\left( \phi ,\mathcal{P}\right) \right) ;
\label{Z-1} \\
\Updownarrow  \notag \\
\mathcal{Z}^{^{\dag }}\equiv -\Xi ^{^{\dag }}\Omega \mathcal{P}=f^{^{\dag
}}q-h^{^{\dag }}p=\left(
\begin{array}{c}
\overline{Z}^{AB}\left( \phi ,\mathcal{P}\right) \\
\\
\overline{Z}^{I}\left( \phi ,\mathcal{P}\right)
\end{array}
\right) ;  \label{Z-2} \\
\notag \\
Z_{AB}\left( \phi ,\mathcal{P}\right) \equiv f_{AB}^{\Lambda }q_{\Lambda
}-h_{AB\mid \Lambda }p^{\Lambda };  \label{Z-3} \\
\notag \\
Z_{I}\left( \phi ,\mathcal{P}\right) \equiv \overline{f}_{I}^{\Lambda
}q_{\Lambda }-\overline{h}_{I\mid \Lambda }p^{\Lambda }.  \label{Z-4}
\end{gather}
Thus, Eq. (\ref{VBH-3}) yields the \textit{``BH potential''} $V_{BH}\left(
\phi ,\mathcal{P}\right) $ to be nothing but the sum of the squares of the
\textit{``dressed''} charges. It is here worth noticing that $\left(
f_{AB}^{\Lambda },\overline{f}_{I}^{\Lambda }\right) $ and $\left( h_{AB\mid
\Lambda },\overline{h}_{I\mid \Lambda }\right) $ are $n\times n$ complex
matrices, because it holds that\footnote{%
Unless otherwise noted, square brackets denote antisymmetrization with
respect to the enclosed indices.} $f_{AB}^{\Lambda }=f_{\left[ AB\right]
}^{\Lambda }$, $h_{AB\mid \Lambda }=h_{\left[ AB\right] \mid \Lambda }$
(thus implying $Z_{AB}=Z_{\left[ AB\right] }$), and
\begin{equation}
n=\frac{\mathcal{N}\left( \mathcal{N}-1\right) }{2}+m,  \label{numbers}
\end{equation}
where $\mathcal{N}$ stands for the number of \textit{spinorial supercharges}
(see Footnote 4), and $m$ denotes the number of \textit{matter multiplets}
coupled to the supergravity multiplet, except for $\mathcal{N}=6$, $d=4$
\textit{pure} supergravity, for which $m=1$.

Eqs. (\ref{Z-3}) and (\ref{Z-4}) are the basic relation between the
(scalar-dependent) \textit{``dressed''} charges $Z_{AB}$ and $Z_{I}$ and the
(scalar-independent) \textit{``bare'' }charges $\mathcal{P}$. It is worth
remarking that $Z_{AB}$ is the \textit{``central charge matrix function'', }%
whose asymptotical value appears in the right-hand side of the $\mathcal{N}$%
-extended ($d=4$) supersymmetry algebra, pertaining to the asymptotical
Minkowski space-time background:
\begin{equation}
\left\{ Q_{\alpha }^{A},Q_{\beta }^{B}\right\} =\epsilon _{\alpha \beta
}Z^{AB}\left( \phi _{\infty },\mathcal{P}\right) ,  \label{susy-algebra}
\end{equation}
where $\phi _{\infty }$ denotes the set of values taken by the scalar fields
at \textit{radial infinity} ($r\rightarrow \infty $) within the considered
static, spherically symmetric and asymptotically flat \textit{dyonic}
\textit{extremal} BH background. Notice that the indices $A$, $B$\ of the
\textit{central charge matrix} are raised and lowered with the metric of the
relevant $\mathcal{R}$\textit{-symmetry }group of the corresponding
supersymmetry algebra.

By denoting the \textit{ADM mass} \cite{ADM}\textbf{\ }of the considered BH
background by $M_{ADM}\left( \phi _{\infty },\mathcal{P}\right) $, the
\textit{BPS bound} \cite{BPS} implies that
\begin{equation}
M_{ADM}\left( \phi _{\infty },\mathcal{P}\right) \geqslant \left| \mathbf{Z}%
_{1}\left( \phi _{\infty },\mathcal{P}\right) \right| \geqslant ...\geqslant
\left| \mathbf{Z}_{\left[ \mathcal{N}/2\right] }\left( \phi _{\infty },%
\mathcal{P}\right) \right| ,  \label{BPS-N}
\end{equation}
where $\mathbf{Z}_{1}\left( \phi ,\mathcal{P}\right) ,...,\mathbf{Z}_{\left[
\mathcal{N}/2\right] }\left( \phi ,\mathcal{P}\right) $ denote the set of
\textit{skew-eigenvalues} of $Z_{AB}\left( \phi ,\mathcal{P}\right) $, and
here square brackets denote the integer part of the enclosed number. If $%
1\leqslant \mathbf{k}\leqslant \left[ \mathcal{N}/2\right] $ of the bounds
expressed by Eq. (\ref{BPS-N}) are saturated, the corresponding extremal BH
state is named to be $\frac{\mathbf{k}}{\mathcal{N}}$-BPS. Thus, the minimal
fraction of total supersymmetries (pertaining to the asymptotically flat
space-time metric) preserved by the extremal BH background within the
considered assumptions is $\frac{1}{\mathcal{N}}$ (for $\mathbf{k}=1$),
while the maximal one is $\frac{1}{2}$ (for $\mathbf{k}=\frac{\mathcal{N}}{2}
$). See Sect. \ref{ADM-Mass} for further details.\medskip

We end the present Section with some considerations on the issue of duality
invariants.

A \textit{duality} invariant $\mathcal{I}$ is a suitable linear combination
(in general with complex coefficients) of ($\phi $-dependent) $H$-invariant
combinations of $Z_{AB}\left( \phi ,\mathcal{P}\right) $ and $Z_{I}\left(
\phi ,\mathcal{P}\right) $ such that Eq. (\ref{U-inv}) holds, \textit{i.e.}
such that $\mathcal{I}$ is invariant under $G$, and thus $\phi $%
-independent:
\begin{equation}
\mathcal{I}=\mathcal{I}\left( Z_{AB}\left( \phi ,\mathcal{P}\right)
,Z_{I}\left( \phi ,\mathcal{P}\right) \right) =\mathcal{I}\left( \mathcal{P}%
\right) .
\end{equation}

In presence of \textit{matter coupling}, a charge configuration $\mathcal{P}$
(and thus a certain orbit of the symplectic representation of the $U$%
-duality group $G$, to which $\mathcal{P}$ belongs) is called \textit{%
supersymmetric} \textit{iff}, by suitably specifying $\phi =\phi \left(
\mathcal{P}\right) $, it holds that
\begin{equation}
Z_{I}\left( \phi \left( \mathcal{P}\right) ,\mathcal{P}\right) =0,~\forall
I=1,...,m.  \label{susy-conds}
\end{equation}

Notice that the conditions (\ref{susy-conds}) cannot hold \textit{identically%
} in $\phi $, otherwise such conditions would be $G$-invariant, which
generally are \textit{not}. Indeed, in order for the supersymmetry
constraints (\ref{susy-conds}) to be invariant (or covariant) under $G$, the
following conditions must hold \textit{identically} in $\phi $:
\begin{equation}
\partial _{\phi }Z_{I}\left( \phi ,\mathcal{P}\right) =0,~~\forall \phi \in
M.
\end{equation}

Therefore, supersymmetry conditions are \textit{not} generally $G$-invariant
(\textit{i.e.} $U$-invariant), otherwise extremal BH attractors (which are
\textit{large}) supported by supersymmetric charge configurations would
not exist.

Nevertheless, in some supergravities it is possible to give $U$-invariant
supersymmetry conditions. In light of previous reasoning, such $U$-invariant
supersymmetric conditions cannot stabilize the scalar fields in terms of
charges (by implementing the \textit{attractor mechanism} in the considered
framework), because such $U$-invariant conditions are actually \textit{%
identities}, \textit{and not equations}, for the set of scalar fields $\phi $%
. Actually, $U$-invariant supersymmetry conditions can be given for all
supersymmetric charge orbits supporting \textit{small} BHs (for which
the classical \textit{attractor mechanism} does not hold). This can be seen
\textit{e.g.} in $\mathcal{N}=8$ (\textit{pure}) and $\mathcal{N}=4$ (%
\textit{matter coupled}) $d=4$ supergravities, respectively treated in
Sects. \ref{N=8-ungauged} and \ref{N=4-ungauged}.

\section{\label{N=8-ungauged}$\mathcal{N}=8$}

The scalar manifold of the maximal, namely $\mathcal{N}=8$, supergravity in $%
d=4$ is the \textit{symmetric} real coset
\begin{equation}
\left( \frac{G}{H}\right) _{\mathcal{N}=8,d=4}=\frac{E_{7\left( 7\right) }}{%
SU\left( 8\right) },~dim_{\mathbb{R}}=70,  \label{N=8-scalar-manifold}
\end{equation}
where the usual notation for non-compact forms of exceptional Lie groups is
used, with subscripts denoting the difference \textit{``}$\#$\textit{\
non-compact generators }$-$\textit{\ }$\#$\textit{\ compact generators''}.
This theory is \textit{pure}, \textit{i.e.} matter coupling is \textit{not}
allowed. The \textit{classical} (see Footnote 1) $U$-duality group is $%
E_{7\left( 7\right) }$. Moreover, the $\mathcal{R}$-symmetry group is $%
SU\left( 8\right) $ and, due to the absence of \textit{matter multiplets},
it is nothing but the stabilizer of the scalar manifold (\ref
{N=8-scalar-manifold}) itself.

The Abelian vector field strengths and their \textit{duals}, as well the
corresponding \textit{fluxes} (charges), sit in the \textit{fundamental}
representation $\mathbf{56}$ of the global, \textit{classical} $U$-duality
group $E_{7\left( 7\right) }$. Such a representation determines the
embedding of $E_{7\left( 7\right) }$ into the symplectic group $Sp\left( 56,%
\mathbb{R}\right) $, which is the largest symmetry acting linearly on
charges. The $\mathbf{56}$ of $E_{7\left( 7\right) }$ admits an \textit{%
unique} invariant, which will be denoted by $\mathcal{I}_{4,\mathcal{N}=8}$
throughout. $\mathcal{I}_{4,\mathcal{N}=8}$ is \textit{quartic} in charges,
and it was firstly determined in \cite{CJ}.

More precisely, $\mathcal{I}_{4,\mathcal{N}=8}$ is the \textit{unique}
combination of $Z_{AB}\left( \phi ,\mathcal{P}\right) $ satisfying
\begin{equation}
\partial _{\phi }\mathcal{I}_{4,\mathcal{N}=8}\left( Z_{AB}\left( \phi ,%
\mathcal{P}\right) \right) =0,~~\forall \phi \in \frac{E_{7\left( 7\right) }%
}{SU\left( 8\right) }.  \label{N=8-G-inv}
\end{equation}
Eq. (\ref{N=8-G-inv}) can be computed by using the \textit{Maurer-Cartan Eqs.%
} of the coset $\frac{E_{7\left( 7\right) }}{SU\left( 8\right) }$ (see
\textit{e.g.} \cite{ADF-U-duality-d=4} and Refs. therein):
\begin{equation}
\nabla Z_{AB}=\frac{1}{2}P_{ABCD}\overline{Z}^{CD},  \label{N=8-MC}
\end{equation}
or equivalently by performing an infinitesimal $\frac{E_{7\left( 7\right) }}{%
SU\left( 8\right) }$-transformation of the central charge matrix (see
\textit{e.g.} \cite{ADF-U-duality-d=4} and Refs. therein):
\begin{equation}
\delta _{\xi _{ABCD}}Z_{AB}=\frac{1}{2}\xi _{ABCD}\overline{Z}^{CD},
\label{N=8-inf-tr}
\end{equation}
where $\nabla $ and $P_{ABCD}$ respectively denote the covariant
differential operator and the \textit{Vielbein} $1$-form in $\frac{%
E_{7\left( 7\right) }}{SU\left( 8\right) }$, and the infinitesimal $\frac{%
E_{7\left( 7\right) }}{SU\left( 8\right) }$-parameters $\xi _{ABCD}$ satisfy
the reality constraint
\begin{equation}
\xi _{ABCD}=\frac{1}{4!}\epsilon _{ABCDEFGH}\overline{\xi }^{EFGH}.
\label{N=8-inf-tr-2}
\end{equation}
As firstly found in \cite{CJ} and rigorously re-obtained in \cite
{ADF-U-duality-d=4}, the unique solution of Eq. (\ref{N=8-G-inv}) reads:
\begin{equation}
\mathcal{I}_{4,\mathcal{N}=8}=\frac{1}{2^{2}}\left[ 2^{2}Tr\left( \left(
Z_{AC}\overline{Z}^{BC}\right) ^{2}\right) -\left( Tr\left( Z_{AC}\overline{Z%
}^{BC}\right) \right) ^{2}+2^{5}Re\left( Pf\left( Z_{AB}\right) \right) %
\right] ,  \label{I4-N=8-Cremmer-Julia}
\end{equation}
where the \textit{Pfaffian} of $Z_{AB}$ is defined as \cite{CJ}
\begin{equation}
Pf\left( Z_{AB}\right) \equiv \frac{1}{2^{4}4!}\epsilon
^{ABCDEFGH}Z_{AB}Z_{CD}Z_{EF}Z_{GH},  \label{N=8-Pfaffian}
\end{equation}
and it holds that (see \textit{e.g.} \cite{ADF-U-duality-d=4})
\begin{equation}
\left| Pf\left( Z_{AB}\right) \right| =\left| det\left( Z_{AB}\right)
\right| ^{1/2}.
\end{equation}
In \cite{ADF-U-duality-d=4} it was indeed shown that, although each of the
three terms of the expression (\ref{I4-N=8-Cremmer-Julia}) is $SU\left(
8\right) $-invariant but \textit{scalar-dependent}, only the combination
given by the expression (\ref{I4-N=8-Cremmer-Julia}) is actually $E_{7\left(
7\right) }$-independent and thus \textit{scalar-independent}, satisfying
\begin{equation}
\delta _{\xi _{ABCD}}\mathcal{I}_{4,\mathcal{N}=8}=0,
\end{equation}
with Eqs. (\ref{N=8-inf-tr}) and (\ref{N=8-inf-tr-2}) holding true.\medskip

It is here worth commenting a bit further about formula (\ref
{I4-N=8-Cremmer-Julia}). The first two terms in its right-hand side are
actually $U\left( 8\right) $-invariant, while the third one, namely $%
2^{5}Re\left( Pf\left( Z_{AB}\right) \right) $, is \textit{only} $SU\left(
8\right) $-invariant. Such a third term introduces an $SU\left( 8\right) $%
-invariant phase $\varphi _{Z}$, defined as (one fourth of) the overall
phase of the central charge matrix, when this latter is reduced to a
skew-diagonal form in the so-called \textit{normal frame} through an $%
SU\left( 8\right) $-transformation:
\begin{equation}
Z_{AB}\overset{SU\left( 8\right) }{\longrightarrow }Z_{AB,skew-diag.}\equiv
e^{i\varphi _{Z}/4}\left(
\begin{array}{cccc}
e_{1} &  &  &  \\
& e_{2} &  &  \\
&  & e_{3} &  \\
&  &  & e_{4}
\end{array}
\right) \otimes \epsilon ,~~e_{i}\in \mathbb{R}^{+},~\forall i=1,...,4,
\label{ZAB-normal-frame}
\end{equation}
where the ordering $e_{1}\geqslant e_{2}\geqslant e_{3}\geqslant e_{4}$ can
be performed without any loss of generality, and the $2\times 2$ symplectic
metric
\begin{equation}
\epsilon \equiv \left(
\begin{array}{cc}
0 & -1 \\
1 & 0
\end{array}
\right)  \label{epsilon}
\end{equation}
has been introduced (notice $\epsilon =\Omega $ for $n=1$, as defined in Eq.
(\ref{Omega})). For non-vanishing (in general all different) \textit{%
skew-eigenvalues} $e_{i}$, the symmetry group of $Z_{AB,skew-diag.}$ is $%
\left( USp\left( 2\right) \right) ^{4}\sim \left( SU\left( 2\right) \right)
^{4}$. Thus, beside the $4$ \textit{skew-eigenvalues} $e_{i}$ and the phase $%
\varphi _{Z}$, the generic $Z_{AB}$ is described by $51=dim_{\mathbb{R}%
}\left( \frac{SU\left( 8\right) }{\left( SU\left( 2\right) \right) ^{4}}%
\right) $ \textit{``generalized angles''}. Consistently, the total number of
parameters is $4+1+51=56$, which is the real dimension of the \textit{%
fundamental} representation $\mathbf{56}$, defining the embedding of $%
E_{7\left( 7\right) }$ into $Sp\left( 56,\mathbb{R}\right) $.

Equivalently, $\varphi _{Z}$ can be defined through the \textit{Pfaffian} of
$Z_{AB}$ as follows:
\begin{equation}
e^{2i\varphi _{Z}}\equiv \frac{Pf\left( Z_{AB}\right) }{Pf\left( \overline{Z}%
_{AB}\right) },
\end{equation}
where clearly $Pf\left( \overline{Z}_{AB}\right) =\overline{Pf\left(
Z_{AB}\right) }$, as yielded by the definition (\ref{N=8-Pfaffian}). It is
then immediate to compute $\varphi _{Z}$ from Eq. (\ref{I4-N=8-Cremmer-Julia}%
):
\begin{equation}
cos\varphi _{Z}\left( \phi ,\mathcal{P}\right) =\frac{\left[ 2^{2}\mathcal{I}%
_{4,\mathcal{N}=8}\left( \mathcal{P}\right) -2^{2}Tr\left( \left( Z_{AC}%
\overline{Z}^{BC}\right) ^{2}\right) +\left( Tr\left( Z_{AB}\overline{Z}%
^{AC}\right) \right) ^{2}\right] }{2^{5}\left( det\left( Z_{AC}\overline{Z}%
^{BC}\right) \right) ^{1/4}}.  \label{N=8-phase}
\end{equation}
Notice that through Eq. (\ref{N=8-phase}) ($cos$)$\varphi _{Z}$ is
determined in terms of the scalar fields $\phi $ and of the BH charges $%
\mathcal{P}$, also along the \textit{small} orbits where $\mathcal{I}_{4,%
\mathcal{N}=8}=0$. However, Eq. (\ref{N=8-phase}) is not defined in the
cases in which $det\left( Z_{AC}\overline{Z}^{BC}\right) =0$, \textit{i.e.}
when \textit{at least} one of the eigenvalues of the matrix $Z_{AC}\overline{%
Z}^{BC}$ vanishes. In such cases, $\varphi _{Z}$ is actually
undetermined.\medskip

In $\mathcal{N}=8$, $d=4$ supergravity five distinct orbits of the $\mathbf{%
56}$ of $E_{7\left( 7\right) }$ exist, as resulting from the analyses
performed in \cite{FG1} and \cite{LPS}. They can be classified in \textit{%
large} and \textit{small }charge orbits, depending whether they
correspond to $\mathcal{I}_{4,\mathcal{N}=8}\neq 0$ or $\mathcal{I}_{4,%
\mathcal{N}=8}=0$, respectively.\smallskip\

Only two \textit{large} charge orbits (for which $\mathcal{I}_{4,%
\mathcal{N}=8}\neq 0$, and the \textit{attractor mechanism} holds) exist in $%
\mathcal{N}=8$, $d=4$ supergravity:

\begin{enumerate}
\item  The \textit{large }$\frac{1}{8}$-BPS orbit \cite{FG1,LPS}
\begin{equation}
\mathcal{O}_{\frac{1}{8}-BPS,\text{\textit{large}}}=\frac{E_{7\left(
7\right) }}{E_{6\left( 2\right) }},\text{~}dim_{\mathbb{R}}=55,
\label{1/8-BPS-large}
\end{equation}
is defined by the $E_{7\left( 7\right) }$-invariant constraint
\begin{equation}
\mathcal{I}_{4,\mathcal{N}=8}>0.
\end{equation}
At the event horizon of the extremal BH, the solution of the $\mathcal{N}=8$%
, $d=4$ \textit{Attractor Eqs.} yields \cite{FM,DFL-0-brane,FK-N=8}
\begin{equation}
e_{1}\in \mathbb{R}_{0}^{+},~~e_{2}=e_{3}=e_{4}=0,  \label{1/8-BPS-large-sol}
\end{equation}
implying $det\left( Z_{AB}\right) =0\Leftrightarrow Pf\left( Z_{AB}\right)
=0 $, and thus $\varphi _{Z}$ to be \textit{undetermined}. Thus, at the
event horizon, the symmetry of the \textit{skew-diagonalized} central charge
matrix $Z_{AB,skew-diag.}$ defined in Eq. (\ref{ZAB-normal-frame}) gets
\textit{enhanced} as follows, revealing the maximal compact symmetry of $%
\mathcal{O}_{\frac{1}{8}-BPS,\text{\textit{large}}}$:
\begin{equation}
\left( USp\left( 2\right) \right) ^{4}\overset{r\rightarrow r_{H}^{+}}{%
\longrightarrow }USp\left( 2\right) \times SU\left( 6\right) \sim SU\left(
2\right) \times SU\left( 6\right) .  \label{1/8-BPS-large-enhancement}
\end{equation}
Indeed, $SU\left( 2\right) \times SU\left( 6\right) $ is the \textit{maximal
compact subgroup} ($mcs$, with symmetric embedding \cite{Gilmore}) of $%
E_{6\left( 2\right) }$ (stabilizer of $\mathcal{O}_{\frac{1}{8}-BPS,\text{%
\textit{large}}}$) itself.

\item  The \textit{large }non-BPS ($Z_{AB}\neq 0$) orbit \cite{FG1,LPS}
\begin{equation}
\mathcal{O}_{non-BPS,Z_{AB}\neq 0}=\frac{E_{7\left( 7\right) }}{E_{6\left(
6\right) }},\text{~}dim_{\mathbb{R}}=55,  \label{non-BPS}
\end{equation}
is defined by the $E_{7\left( 7\right) }$-invariant constraint
\begin{equation}
\mathcal{I}_{4,\mathcal{N}=8}<0.
\end{equation}
At the event horizon of the extremal BH, the solution of the $\mathcal{N}=8$%
, $d=4$ \textit{Attractor Eqs.} yields \cite{FM,DFL-0-brane,FK-N=8}
\begin{equation}
e_{1}=e_{2}=e_{3}=e_{4}\in \mathbb{R}_{0}^{+},~~\varphi _{Z}=\pi +2k\pi
,~k\in \mathbb{Z,}  \label{non-BPS-sol}
\end{equation}
so the \textit{skew-eigenvalues} of $Z_{AB}$\ at the horizon (see Eq. (\ref
{ZAB-normal-frame})) are complex. Thus, at the event horizon, the symmetry
of the \textit{skew-diagonalized} central charge matrix $Z_{AB,skew-diag.}$
defined in Eq. (\ref{ZAB-normal-frame}) gets \textit{enhanced} as follows,
revealing the maximal compact symmetry of $\mathcal{O}_{non-BPS,Z_{AB}\neq
0} $:
\begin{equation}
\left( USp\left( 2\right) \right) ^{4}\overset{r\rightarrow r_{H}^{+}}{%
\longrightarrow }USp\left( 8\right) .  \label{non-BPS-enhancement}
\end{equation}
Indeed, $USp\left( 8\right) $ is the $mcs$ (with symmetric embedding \cite
{Gilmore}) of $E_{6\left( 6\right) }$ (stabilizer of $\mathcal{O}%
_{non-BPS,Z_{AB}\neq 0}$) itself.\bigskip
\end{enumerate}

As mentioned above, for such \textit{large }charge orbits, corresponding
to a non-vanishing quartic $E_{7\left( 7\right) }$-invariant $\mathcal{I}_{4,%
\mathcal{N}=8}$ and thus supporting \textit{large} BHs, the \textit{%
attractor mechanism} holds. Consequently, the computations of the
Bekenstein-Hawking BH entropy can be performed by solving the criticality
conditions for the \textit{``BH potential''}
\begin{equation}
V_{BH,\mathcal{N}=8}=\frac{1}{2}Z_{AB}\overline{Z}^{AB},  \label{N=8-VBH}
\end{equation}
the result being
\begin{equation}
\frac{S_{BH,}}{\pi }=\left. V_{BH,\mathcal{N}=8}\right| _{\partial V_{BH,%
\mathcal{N}=8}=0}=V_{BH,\mathcal{N}=8}\left( \phi _{H}\left( \mathcal{P}%
\right) ,\mathcal{P}\right) =\left| \mathcal{I}_{4,\mathcal{N}=8}\right|
^{1/2},  \label{N=8-AM}
\end{equation}
where $\phi _{H}\left( \mathcal{P}\right) $ denotes the set of solutions to
the \textit{criticality conditions} of $V_{BH,\mathcal{N}=8}$, namely the
\textit{Attractor Eqs.} of $\mathcal{N}=8$, $d=4$ supergravity:
\begin{equation}
\partial _{\phi }V_{BH,\mathcal{N}=8}=0,~~\forall \phi \in \frac{E_{7\left(
7\right) }}{SU\left( 8\right) },  \label{N=8-AEs}
\end{equation}
expressing the stabilization of the scalar fields purely in terms of
supporting charges $\mathcal{P}$ at the event horizon of the extremal BH.
Through Eqs. (\ref{N=8-MC}) and (\ref{N=8-VBH}), Eqs. (\ref{N=8-AEs}) can be
rewritten as follows (notice the strict similarity to Eq. (\ref
{N=8-doubly-critical-2}) further below) \cite{FK-N=8}:
\begin{equation}
Z_{[AB}Z_{CD]}+\frac{1}{4!}\epsilon _{ABCDEFGH}\overline{Z}^{EF}\overline{Z}%
^{GH}=0.  \label{N=8-AEs-2}
\end{equation}

Actually, the critical potential $\left. V_{BH,\mathcal{N}=8}\right|
_{\partial V_{BH,\mathcal{N}=8}=0}$ exhibits some \textit{``flat''}
directions, so not all scalars are stabilized in terms of charges at the
event horizon \cite{Ferrara-Marrani-1,Ferrara-Marrani-2}. Thus, Eq. (\ref
{N=8-AM}) yields that the \textit{unstabilized} scalars, spanning a related
\textit{moduli space} of the considered class of attractor solutions, do not
enter in the expression of the BH entropy at all. The \textit{moduli spaces}%
\footnote{%
Results obtained by explicit computations within the $\mathcal{N}=2$, $d=4$
\textit{symmetric} so-called $stu$ model in \cite{GLS} and \cite
{stu-unveiled} seem to point out that the \textit{moduli spaces} should be
present not only at the event horizon of the considered extremal BH (\textit{%
i.e.} for $r\rightarrow r_{H}^{+}$), but also all along the scalar attractor
\textit{flow} (\textit{i.e.} $\forall r\geqslant r_{H}$).} exhibited by the
\textit{Attractor Eqs.} (\ref{N=8-AEs})-(\ref{N=8-AEs-2}) are \cite
{Ferrara-Marrani-2}
\begin{eqnarray}
\mathcal{M}_{\frac{1}{8}-BPS,\text{\textit{large}}} &=&\frac{E_{6\left(
2\right) }}{SU\left( 2\right) \times SU\left( 6\right) },~dim_{\mathbb{R}%
}=40; \\
\mathcal{M}_{non-BPS,Z_{AB}\neq 0} &=&\frac{E_{6\left( 6\right) }}{USp\left(
8\right) },~dim_{\mathbb{R}}=42.  \label{Stanford-1}
\end{eqnarray}
As found in \cite{Ferrara-Marrani-2}, the general structure of the \textit{%
moduli spaces} of attractor solutions in supergravities based on \textit{%
symmetric} scalar manifolds $\frac{G}{H}$ is
\begin{equation}
\frac{\mathcal{H}_{nc}}{\mathbf{h}},
\end{equation}
where $\mathcal{H}_{nc}$ is the \textit{non-compact} stabilizer of the
charge orbit $\frac{G}{\mathcal{H}_{nc}}$ (apart from eventual $U\left(
1\right) $ factors, $\mathcal{H}_{nc}$ is a non-compact, real form of $H$),
and $\mathbf{h}=mcs\left( \mathcal{H}_{nc}\right) $. As justified in \cite
{ADF-U-duality-d=4} and then in \cite{Ferrara-Marrani-1}, $\mathcal{M}_{%
\frac{1}{8}-BPS,\text{\textit{large}}}$ is a \textit{quaternionic} symmetric
manifold. Furthermore, $\mathcal{M}_{non-BPS,Z_{AB}\neq 0}$ given by Eq. (%
\ref{Stanford-1}) is nothing but the scalar manifold of $\mathcal{N}=8$, $%
d=5 $ supergravity. The stabilizers of $\mathcal{M}_{\frac{1}{8}-BPS,\text{%
\textit{large}}}$ and $\mathcal{M}_{non-BPS,Z_{AB}\neq 0}$ exploit the
maximal compact symmetry of the corresponding charge orbits; this symmetry
becomes fully manifest through the enhancement of the compact symmetry group
of $Z_{AB,skew-diag.}$ at the event horizon of the extremal BH, respectively
given by Eqs. (\ref{1/8-BPS-large-enhancement}) and (\ref
{non-BPS-enhancement}). \smallskip\

It is now convenient to denote with $\lambda _{i}$ ($i=1,...,4$) the four
real non-negative eigenvalues of the matrix $Z_{AB}\overline{Z}^{CB}=\left(
ZZ^{\dag }\right) _{A}^{C}$. By recalling Eq. (\ref{ZAB-normal-frame}), one
can notice that
\begin{equation}
\lambda _{i}=e_{i}^{2},  \label{lambda-e}
\end{equation}
and one can order them as $\lambda _{1}\geqslant \lambda _{2}\geqslant
\lambda _{3}\geqslant \lambda _{4}$, without any loss of
generality.\smallskip\ The explicit expression of $\lambda _{i}$ in terms of
$U\left( 8\right) $-invariants (namely of $Tr\left( ZZ^{\dag }\right) $, $%
Tr\left( \left( ZZ^{\dag }\right) ^{2}\right) $, $Tr\left( \left( ZZ^{\dag
}\right) ^{3}\right) $ and $Tr\left( \left( ZZ^{\dag }\right) ^{4}\right) $,
and suitable powers) is given by Eqs. (4.74), (4.75), (4.86) and (4.87) of
\cite{DFL-0-brane}, and it will be used in Sect. \ref{ADM-Mass} to determine
the \textit{ADM mass} for $\frac{\mathbf{k}}{8}$-BPS ($\mathbf{k}=1,2,4$)
extremal BH states. \

Three distinct \textit{small} charge orbits (all with $\mathcal{I}_{4,%
\mathcal{N}=8}=0$) exist, and they all are supersymmetric :

\begin{enumerate}
\item  The generic \textit{small lightlike} orbit is $\frac{1}{8}$-BPS,
it is defined by the $E_{7\left( 7\right) }$-invariant constraint
\begin{equation}
\mathcal{I}_{4,\mathcal{N}=8}=0,  \label{N=8-lightlike}
\end{equation}
and it reads \cite{FG1,LPS}
\begin{equation}
\mathcal{O}_{\frac{1}{8}-BPS,\text{\textit{small}}}=\frac{E_{7\left(
7\right) }}{F_{4\left( 4\right) }\times _{s}T_{26}},~dim_{\mathbb{R}}=55.
\label{1/8-BPS-small}
\end{equation}
Generally, it yields four different $\lambda _{i}$'s, and in this case Eq. (%
\ref{N=8-phase}) reduces to
\begin{equation}
\left. cos\varphi _{Z}\left( \phi ,\mathcal{P}\right) \right| _{\mathcal{I}%
_{4,\mathcal{N}=8}=0}=\left. -\frac{\left[ 2^{2}Tr\left( \left( Z_{AC}%
\overline{Z}^{BC}\right) ^{2}\right) -\left( Tr\left( Z_{AB}\overline{Z}%
^{AC}\right) \right) ^{2}\right] }{2^{5}\left( det\left( Z_{AC}\overline{Z}%
^{BC}\right) \right) ^{1/4}}\right| _{\mathcal{I}_{4,\mathcal{N}=8}=0}.
\end{equation}
In agreement with the results of \cite{FG1} and \cite{LPS}, the (maximal
compact) symmetry of the \textit{skew-diagonalized} central charge matrix $%
Z_{AB,skew-diag.}$ all along the $\frac{1}{8}$-BPS \textit{small} flow
is the generic one: $\left( SU\left( 2\right) \right) ^{4}$. The counting of
the parameters of $\mathcal{O}_{\frac{1}{8}-BPS,\text{\textit{small}}}$
consistently reads: $55=4$ \textit{skew-eigenvalues} $\lambda _{i}+1$ phase $%
\varphi _{Z}+51\left( =dim_{\mathbb{R}}\left( \frac{SU\left( 8\right) }{%
\left( SU\left( 2\right) \right) ^{4}}\right) \right) $ \textit{%
``generalized angles''}$-1$ defining constraint (\ref{N=8-lightlike}).

\item  The \textit{small critical }orbit is $\frac{1}{4}$-BPS. It reads%
\textbf{\ } \cite{FG1,LPS}
\begin{equation}
\mathcal{O}_{\frac{1}{4}-BPS}=\frac{E_{7\left( 7\right) }}{\left(
SO(6,5)\times _{s}T_{32}\right) \times T_{1}},~dim_{\mathbb{R}}=45,
\label{1/4-BPS}
\end{equation}
and it is defined by the following differential constraint on $\mathcal{I}%
_{4,\mathcal{N}=8}$ \cite{FM,DFL-0-brane}:
\begin{equation}
\frac{\partial \mathcal{I}_{4,\mathcal{N}=8}}{\partial Z_{AB}}=0,
\label{N=8-critical}
\end{equation}
which, due to the reality of $\mathcal{I}_{4,\mathcal{N}=8}$, is actually $%
E_{7\left( 7\right) }$-invariant. Let us also notice that, due to the
homogeneity of $\mathcal{I}_{4,\mathcal{N}=8}$ of degree four in $\mathcal{P}
$, Eq. (\ref{N=8-critical}) implies the constraint (\ref{N=8-lightlike}). In
particular, along the $\frac{1}{4}$-BPS orbit it holds that (the labelling
does not yield any loss of generality)
\begin{equation}
\lambda _{1}=\lambda _{2}>\lambda _{3}=\lambda _{4}\geqslant 0.
\label{N=8-1/4-BPS-structure}
\end{equation}
If $Pf\left( Z_{AB}\right) \neq 0$ then
\begin{equation}
\lambda _{1}=\lambda _{2}>\lambda _{3}=\lambda _{4}>0,
\label{N=8-critical-2-pre}
\end{equation}
and Eq. (\ref{N=8-phase}) yields $\varphi _{Z}=k\pi $, $k\in \mathbb{Z}$, so
the \textit{skew-eigenvalues} of $Z_{AB}$ (see Eq. (\ref{ZAB-normal-frame}))
are real and the (maximal) compact symmetry of $Z_{AB,skew-diag.}$ is $%
\left( USp\left( 4\right) \right) ^{2}$. On the other hand, if $Pf\left(
Z_{AB}\right) =0$ then
\begin{equation}
\lambda _{1}=\lambda _{2}>\lambda _{3}=\lambda _{4}=0,
\label{N=8-critical-2}
\end{equation}
and $\varphi _{Z}$ is undetermined. In this case, the (maximal compact)
symmetry of the \textit{skew-diagonalized} central charge matrix $%
Z_{AB,skew-diag.}$ is $USp\left( 4\right) \times SU\left( 4\right) \sim
SO\left( 5\right) \times SO\left( 6\right) $, which is the $mcs$ of the
non-translational part of the stabilizer of $\mathcal{O}_{\frac{1}{4}-BPS}$,
expressing the maximal compact symmetry of $\mathcal{O}_{\frac{1}{4}-BPS}$
itself. In agreement with the results of \cite{FG1} and \cite{LPS}, the
\textit{maximal} (compact) symmetry of the \textit{skew-diagonalized}
central charge matrix $Z_{AB,skew-diag.}$ along the $\frac{1}{4}$-BPS
\textit{small} flow (fully manifest in the particular solution (\ref
{N=8-critical-2})) is $USp\left( 4\right) \times SU\left( 4\right) $. The
counting of the parameters of $\mathcal{O}_{\frac{1}{4}-BPS}$ consistently
reads: $45=2$ \textit{skew-eigenvalues} $\lambda _{1}$ and $\lambda
_{2}+43\left( =dim_{\mathbb{R}}\left( \frac{SU\left( 8\right) }{\left(
USp\left( 4\right) \right) ^{2}}\right) \right) $ \textit{``generalized
angles''}.

\item  The \textit{small doubly-critical }orbit is $\frac{1}{2}$-BPS,
and it reads \cite{FG1,LPS}
\begin{equation}
\mathcal{O}_{\frac{1}{2}-BPS}=\frac{E_{7\left( 7\right) }}{E_{6\left(
6\right) }\times _{s}T_{27}},~dim_{\mathbb{R}}=28.  \label{1/2-BPS}
\end{equation}
It can be defined in an $E_{7\left( 7\right) }$-invariant way by performing
the following two-step procedure \cite{DFL-0-brane}. One starts by
considering the requirement that the second derivative of $\mathcal{I}_{4,%
\mathcal{N}=8}$ (with respect to $Z_{AB}$) projected along the adjoint
representation $\mathbf{Adj}\left( SU\left( 8\right) \right) =\mathbf{63}$
of $SU\left( 8\right) $ vanishes, yielding \cite{DFL-0-brane}
\begin{equation}
\left. \frac{\partial ^{2}\mathcal{I}_{4,\mathcal{N}=8}}{\partial Z_{AB}%
\overline{\partial }\overline{Z}^{BC}}\right| _{\mathbf{Adj}\left( SU\left(
8\right) \right) }=0\Longleftrightarrow Z_{AC}\overline{Z}^{BC}=\frac{1}{%
2^{3}}\delta _{A}^{B}Z_{DE}Z^{DE}.  \label{N=8-doubly-critical-1}
\end{equation}
This is a mixed rank-$2$ $SU\left( 8\right) $-covariant condition. By
further differentiating with respect to the scalars $\phi $ parametrizing $%
\frac{E_{7\left( 7\right) }}{SU\left( 8\right) }$ and using the
Maurer-Cartan Eqs. (\ref{N=8-MC}), one obtains another $SU\left( 8\right) $%
-covariant relation (notice the strict similarity to the $\mathcal{N}=8$, $%
d=4$ \textit{Attractor Eqs.} (\ref{N=8-AEs-2})) \cite{DFL-0-brane}:
\begin{equation}
Z_{[AB}Z_{CD]}-\frac{1}{4!}\epsilon _{ABCDEFGH}\overline{Z}^{EF}\overline{Z}%
^{GH}=0.  \label{N=8-doubly-critical-2}
\end{equation}
Actually, Eq. (\ref{N=8-doubly-critical-2}) form with Eq. (\ref
{N=8-doubly-critical-1}) an $E_{7\left( 7\right) }$-invariant set of
differential conditions defining $\mathcal{O}_{\frac{1}{2}-BPS}$. Indeed, as
noticed in \cite{DFL-0-brane}, Eq. (\ref{N=8-doubly-critical-2}) can be
rewritten as
\begin{equation}
\frac{\partial ^{2}\mathcal{I}_{4,\mathcal{N}=8}}{\partial Z_{[AB}\partial
Z_{CD]}}-\frac{1}{4!}\epsilon ^{ABCDEFGH}\frac{\partial ^{2}\mathcal{I}_{4,%
\mathcal{N}=8}}{\overline{\partial }\overline{Z}^{[EF}\overline{\partial }%
\overline{Z}^{GH]}}=0.  \label{N=8-doubly-critical-3}
\end{equation}
Thus, by using the notation $Z_{\mathbf{56}}\equiv \left( \mathcal{Z},%
\mathcal{Z}^{T}\right) =\left( Z_{AB},\overline{Z}^{AB}\right) $ (recall
Eqs. (\ref{Z-1}) and (\ref{Z-2})), Eqs. (\ref{N=8-doubly-critical-1}) and (%
\ref{N=8-doubly-critical-2})-(\ref{N=8-doubly-critical-3}) can be rewritten
in the manifestly $E_{7\left( 7\right) }$-invariant fashion
\begin{equation}
\left. \frac{\partial ^{2}\mathcal{I}_{4,\mathcal{N}=8}}{\partial Z_{\mathbf{%
56}}\partial Z_{\mathbf{56}}}\right| _{\mathbf{Adj}\left( E_{7\left(
7\right) }\right) }=0,  \label{N=8-doubly-critical-4}
\end{equation}
where $\mathbf{Adj}\left( E_{7\left( 7\right) }\right) =\mathbf{133}$ is the
adjoint representation of $E_{7\left( 7\right) }$. Notice that $\frac{%
\partial ^{2}\mathcal{I}_{4,\mathcal{N}=8}}{\partial Z_{\mathbf{56}}\partial
Z_{\mathbf{56}}}$ is a rank-$2$ symmetric true-tensor $E_{7\left( 7\right) }$%
-tensor, thus sitting in the symmetric product representation $\left(
\mathbf{56}\times \mathbf{56}\right) _{s}=\mathbf{1596}$ of $E_{7\left(
7\right) }$, which in turns enjoys the following branching with respect to $%
E_{7\left( 7\right) }$ \cite{Gilmore, DFL-0-brane}:
\begin{equation}
\left( \mathbf{56}\times \mathbf{56}\right) _{s}=\mathbf{1596}%
\longrightarrow \mathbf{1463}+\underset{\mathbf{Adj}\left( E_{7\left(
7\right) }\right) }{\mathbf{133}}.
\end{equation}
It is here worth remarking that the constraints (\ref{N=8-doubly-critical-1}%
) and (\ref{N=8-doubly-critical-2})-(\ref{N=8-doubly-critical-3}) (or
equivalently ((\ref{N=8-doubly-critical-4}))) imply the constraint (\ref
{N=8-critical}), because in fact they are stronger constraints.
\end{enumerate}

Along the $\frac{1}{2}$-BPS orbit it holds that
\begin{equation}
\lambda _{1}=\lambda _{2}=\lambda _{3}=\lambda _{4}.
\end{equation}
Furthermore, it can be shown that $\varphi _{Z}=2k\pi $, $k\in \mathbb{Z}$,
so the \textit{skew-eigenvalues} of $Z_{AB}$ (see Eq. (\ref{ZAB-normal-frame}%
)) are real. In agreement with the results of \cite{FG1} and \cite{LPS}, the
(maximal compact) symmetry of the \textit{skew-diagonalized} central charge
matrix $Z_{AB,skew-diag.}$ all along the $\frac{1}{2}$-BPS \textit{small}
flow is $USp\left( 8\right) $, which is the $mcs$ of the non-translational
part of the stabilizer of $\mathcal{O}_{\frac{1}{2}-BPS}$, expressing the
maximal compact symmetry of $\mathcal{O}_{\frac{1}{2}-BPS}$ itself. The
counting of the parameters of $\mathcal{O}_{\frac{1}{2}-BPS}$ consistently
reads: $28=1$ \textit{skew-eigenvalue} $\lambda _{1}+27\left( =dim_{\mathbb{R%
}}\left( \frac{SU\left( 8\right) }{USp\left( 8\right) }\right) \right) $
\textit{``generalized angles''}.

Interestingly, $USp\left( 8\right) $ also is the \textit{enhanced} compact
symmetry of $Z_{AB,skew-diag.}$ at the event horizon of the \textit{large%
} non-BPS $Z_{AB}\neq 0$ \textit{attractor} scalar flow (see Eq. (\ref
{non-BPS-enhancement}) above). Indeed, the charge orbits $\mathcal{O}%
_{non-BPS,Z_{AB}\neq 0}$ and $\mathcal{O}_{\frac{1}{2}-BPS}$ (respectively
given by Eqs. (\ref{non-BPS}) and (\ref{1/2-BPS})) coincide, up to the
translational factor $T_{27}$ in the stabilizer, and thus they have the same
maximal compact symmetry.\bigskip\

As given by the analysis of \cite{FM}, the classification of \textit{%
large} and \textit{small} orbits of the $\mathbf{56}$ of $E_{7\left(
7\right) }$ can be performed also considering the symplectic basis composed
by the fluxes $q_{\Lambda }$ ($\Lambda =1,...,56$). In general, the
symplectic basis of charges is useful in order to determine, through
constraints imposed on the relevant $U$-invariant, the number and typology
of orbits of the relevant representation of the $U$-duality group. On the
other hand, using the manifestly $H$-covariant basis of central charges and
matter charges one can achieve a symplectic-invariant characterization of
charge orbits, and also study the related supersymmetry-preserving features.

Finally, it is worth pointing out once again that there is a crucial
difference among the various constraints defining the two \textit{large}
and the three \textit{small} charge orbits of $\mathcal{N}=8$, $d=4$
supergravity listed above:

\begin{itemize}
\item  The \textit{large }charge orbits $\mathcal{O}_{\frac{1}{8}-BPS,%
\text{\textit{large}}}$ and $\mathcal{O}_{non-BPS,Z_{AB}\neq 0}$,
respectively given by Eqs. (\ref{1/8-BPS-large}) and (\ref{non-BPS}), are in
order defined by the $E_{7\left( 7\right) }$-invariant conditions $\mathcal{I%
}_{4,\mathcal{N}=8}>0$ and $\mathcal{I}_{4,\mathcal{N}=8}<0$. Due to their $%
E_{7\left( 7\right) }$-invariance, these conditions are \textit{identities}
for the scalar fields $\phi $ spanning $\frac{E_{7\left( 7\right) }}{%
SU\left( 8\right) }$. However, the \textit{classical attractor mechanism}
does hold for \textit{large} \textit{extremal} BHs, and the scalars $%
\phi $ are stabilized purely in terms of charges $\mathcal{P}$ at the event
horizon ($r\rightarrow r_{H}^{+}$) through the only two independent
solutions (\ref{1/8-BPS-large-sol}) and (\ref{non-BPS-sol}) to the $\mathcal{%
N}=8$, $d=4$ \textit{Attractor Eqs. }(\ref{N=8-AEs})-(\ref{N=8-AEs-2}).

\item  The \textit{small }charge orbits $\mathcal{O}_{\frac{1}{8}-BPS,%
\text{\textit{small}}}$, $\mathcal{O}_{\frac{1}{4}-BPS}$ and $\mathcal{O}_{%
\frac{1}{2}-BPS}$, respectively given by Eqs. (\ref{1/8-BPS-small}), (\ref
{1/4-BPS}) and (\ref{1/2-BPS}), are in order defined by the $E_{7\left(
7\right) }$-invariant conditions (\ref{N=8-lightlike}), (\ref{N=8-critical})
and (\ref{N=8-doubly-critical-4}). Due to their $E_{7\left( 7\right) }$%
-invariance, these conditions are \textit{identities} for the scalars $\phi $%
, which thus are \textit{not} stabilized along such orbits. Indeed, the
\textit{classical attractor mechanism} does \textit{not} hold for \textit{%
small} BHs\textit{.}
\end{itemize}

\section{\label{N=4-ungauged}$\mathcal{N}=4$}

In $\mathcal{N}=4$, $d=4$ supergravity, unlike the $\mathcal{N}=8$ case,
matter (\textit{vector}) multiplets appear (see \textit{e.g.} \cite
{Bergshoeff, De Roo}). By denoting their number with $M$, the related scalar
manifold is the \textit{symmetric} coset
\begin{equation}
\left( \frac{G}{H}\right) _{\mathcal{N}=4,d=4}=\frac{SL\left( 2,\mathbb{R}%
\right) }{U(1)}\times \frac{SO\left( 6,M\right) }{SO\left( 6\right) \times
SO\left( M\right) },~dim_{\mathbb{R}}=6M+2.  \label{N=4-scalar-manifold}
\end{equation}

The Abelian vector field strengths and their \textit{duals}, as well the
corresponding \textit{fluxes} (charges), sit in the bi-fundamental $\left(
\mathbf{2},\mathbf{6+M}\right) $ representation of the global, \textit{%
classical} (see Footnote 1) $U$-duality group $SL\left( 2,\mathbb{R}\right)
\times SO\left( 6,M\right) $ \cite{Cremmer-F-S-1}. Such a representation
determines the embedding of $SL\left( 2,\mathbb{R}\right) \times SO\left(
6,M\right) $ into the symplectic group $Sp\left( 12+2M,\mathbb{R}\right) $.
The representation $\left( \mathbf{2},\mathbf{6+M}\right) $ is endowed with
a natural symplectic metric
\begin{equation}
\mathbf{\Omega }\equiv \epsilon _{\alpha \beta }\eta _{\Lambda \Sigma },
\end{equation}
where $\epsilon _{\alpha \beta }$ ($\alpha ,\beta =1,2$) is the (inverse of
the) $SL\left( 2,\mathbb{R}\right) $ skew-symmetric metric defined in Eq. (%
\ref{epsilon}), and $\eta _{\Lambda \Sigma }$ ($\Lambda ,\Sigma =1,...,6+M=n$%
; recall Eq. (\ref{numbers})) is the Lorentzian metric of $SO\left(
6,M\right) $. Moreover, the $\mathcal{R}$-symmetry group is $U\left(
4\right) $.

Furthermore, $\left( \mathbf{2},\mathbf{6+M}\right) $ admits an \textit{%
unique} invariant, which will be denoted by $\mathcal{I}_{4,\mathcal{N}=4}$
throughout. $\mathcal{I}_{4,\mathcal{N}=4}$ is \textit{quartic} in charges,
and it was firstly determined in \cite{CY,FK2,N=4-pure-BH-entropy}.

More precisely, $\mathcal{I}_{4,\mathcal{N}=4}$ is the \textit{unique}
combination of \textit{``dressed''} charges $Z_{AB}=Z_{[AB]}\left( \phi ,%
\mathcal{P}\right) $ (\textit{central charge matrix}, $A,B=1,...,4$) and $%
Z_{I}\left( \phi ,\mathcal{P}\right) $ (\textit{matter charges}, $I=1,...,M$%
) satisfying
\begin{equation}
\partial _{\phi }\mathcal{I}_{4,\mathcal{N}=4}\left( Z_{AB}\left( \phi ,%
\mathcal{P}\right) ,Z_{I}\left( \phi ,\mathcal{P}\right) \right)
=0,~~\forall \phi \in \left( \frac{G}{H}\right) _{\mathcal{N}=4,d=4}.
\label{N=4-G-inv}
\end{equation}
Eq. (\ref{N=4-G-inv}) can be computed by using the \textit{Maurer-Cartan Eqs.%
} of the coset $\frac{SL\left( 2,\mathbb{R}\right) }{U(1)}\times \frac{%
SO\left( 6,M\right) }{SO\left( 6\right) \times SO\left( M\right) }$ (see
\textit{e.g.} \cite{ADF-U-duality-d=4}, and Refs. therein):
\begin{eqnarray}
\nabla Z_{AB} &=&\frac{1}{2}P\epsilon _{ABCD}\overline{Z}^{CD}+P_{ABI}%
\overline{Z}^{I};  \label{N=4-MC-1} \\
\nabla Z_{I} &=&\frac{1}{2}P_{ABI}\overline{Z}^{AB}+P\eta _{IJ}\overline{Z}%
^{J},  \label{N=4-MC-2}
\end{eqnarray}
or equivalently by performing an infinitesimal $\frac{SL\left( 2,\mathbb{R}%
\right) }{U(1)}\times \frac{SO\left( 6,M\right) }{SO\left( 6\right) \times
SO\left( M\right) }$-transformation of the central charge matrix and of
matter charges (see \textit{e.g.} \cite{ADF-U-duality-d=4}, and Refs.
therein):
\begin{eqnarray}
\delta _{\left( \xi ,\xi _{AB\mid I}\right) }Z_{AB} &=&\frac{1}{2}\xi
\epsilon _{ABCD}\overline{Z}^{CD}+\xi _{AB\mid I}Z^{I};  \label{N=4-inf-tr-1}
\\
\delta _{\left( \xi ,\xi _{AB\mid I}\right) }Z_{I} &=&\overline{\xi }\eta
_{IJ}\overline{Z}^{J}+\frac{1}{2}\xi _{AB\mid I}\overline{Z}^{AB},
\label{N=4-inf-tr-2}
\end{eqnarray}
where $\nabla $ stands for the covariant differential operator in $\frac{%
SL\left( 2,\mathbb{R}\right) }{U(1)}\times \frac{SO\left( 6,M\right) }{%
SO\left( 6\right) \times SO\left( M\right) }$. $P$ and $P_{ABI}$
respectively are the \textit{Vielbein} $1$-forms of $\frac{SL\left( 2,%
\mathbb{R}\right) }{U(1)}$ and $\frac{SO\left( 6,M\right) }{SO\left(
6\right) \times SO\left( M\right) }$, with $P_{ABI}$ satisfying the reality
condition:
\begin{equation}
P_{ABI}=\frac{1}{2}\eta _{IJ}\epsilon _{ABCD}\overline{P}^{CDJ}.
\end{equation}
Moreover, $\xi $ is the infinitesimal $\frac{SL\left( 2,\mathbb{R}\right) }{%
U(1)}$-parameter and $\xi _{AB\mid I}$ are the infinitesimal $\frac{SO\left(
6,M\right) }{SO\left( 6\right) \times SO\left( M\right) }$-parameters,
satisfying the reality condition
\begin{equation}
\overline{\xi }^{AB\mid I}=\frac{1}{2}\eta ^{IJ}\epsilon ^{ABCD}\xi _{CD\mid
J}.  \label{N=4-inf-tr-3}
\end{equation}
As found in \cite{CY,FK2,N=4-pure-BH-entropy} and rigorously re-obtained in
\cite{ADF-U-duality-d=4}, in terms of $Z_{AB}$ and $Z_{I}$ the unique
solution of Eq. (\ref{N=4-G-inv}) reads:
\begin{equation}
\mathcal{I}_{4,\mathcal{N}=4}=\mathcal{S}_{1}^{2}-\left| \mathcal{S}%
_{2}\right| ^{2},  \label{I4-N=4-Z}
\end{equation}
where one can identify $\mathcal{S}_{1}\equiv L_{0}$, $\mathcal{S}%
_{2}=L_{1}+iL_{2}$, with $L\equiv \left( L_{0},L_{1},L_{2}\right) $ being an
$SL\left( 2,\mathbb{R}\right) \sim SO\left( 1,2\right) $-vector with square
norm
\begin{equation}
L^{2}=L_{0}^{2}-L_{1}^{2}-L_{2}^{2}=\mathcal{S}_{1}^{2}-\left| \mathcal{S}%
_{2}\right| ^{2}.  \label{N=4-dressed}
\end{equation}
$\mathcal{S}_{1}$ and $\mathcal{S}_{2}$ are defined as \cite
{ADF-U-duality-d=4}
\begin{eqnarray}
\mathcal{S}_{1} &\equiv &\frac{1}{2}Z_{AB}\overline{Z}^{AB}-Z_{I}\overline{Z}%
^{I}\in \mathbb{R};  \label{S1} \\
\mathcal{S}_{2} &\equiv &\frac{1}{4}\epsilon ^{ABCD}Z_{AB}Z_{CD}-\overline{Z}%
_{I}\overline{Z}_{I}\in \mathbb{C}.  \label{S2}
\end{eqnarray}
In \cite{ADF-U-duality-d=4} it was indeed shown that $\mathcal{I}_{4,%
\mathcal{N}=4}$ given by Eq. (\ref{I4-N=4-Z}) is the unique combination of $%
SO\left( 6,M\right) $-invariant and scalar-dependent quantities, which is
actually \textit{also} $SL\left( 2,\mathbb{R}\right) $-independent and thus
\textit{scalar-independent}, satisfying
\begin{eqnarray}
\delta _{\xi }\mathcal{I}_{4,\mathcal{N}=4} &=&0; \\
\delta _{\xi _{AB\mid I}}\mathcal{I}_{4,\mathcal{N}=4} &=&0,
\end{eqnarray}
with Eqs. (\ref{N=4-inf-tr-1}), (\ref{N=4-inf-tr-2}) and (\ref{N=4-inf-tr-3}%
) holding true.

On the other hand, the expression of $\mathcal{I}_{4,\mathcal{N}=4}$ in
terms of the \textit{``bare''} charges $\mathcal{P}$ reads \cite
{CY,DLR,FK1,FK2}
\begin{equation}
\mathcal{I}_{4,\mathcal{N}=4}=p^{2}q^{2}-\left( p\cdot q\right) ^{2}=\frac{1%
}{2}\left( p_{\Lambda }q_{\Sigma }-p_{\Sigma }q_{\Lambda }\right) \left(
p_{\Xi }q_{\Omega }-p_{\Omega }q_{\Xi }\right) \eta ^{\Lambda \Xi }\eta
^{\Sigma \Omega }=\frac{1}{2}T_{\Lambda \Sigma }^{\left( a\right) }T^{\left(
a\right) \mid \Lambda \Sigma },  \label{N=4-bare}
\end{equation}
where
\begin{equation}
p^{2}\equiv p\cdot p\equiv p_{\Lambda }p_{\Sigma }\eta ^{\Lambda \Sigma
},~q^{2}\equiv q\cdot q\equiv q_{\Lambda }q_{\Sigma }\eta ^{\Lambda \Sigma
},~p\cdot q\equiv p_{\Lambda }q_{\Sigma }\eta ^{\Lambda \Sigma },
\end{equation}
and the tensor
\begin{equation}
T_{\Lambda \Sigma }^{\left( a\right) }\equiv p_{\Lambda }q_{\Sigma
}-p_{\Sigma }q_{\Lambda }=T_{[\Lambda \Sigma ]}^{\left( a\right) }
\label{T-tensor-antisymm}
\end{equation}
has been introduced (the upperscript ``$\left( a\right) $'' stands for
\textit{``anti-symmetric''}).\medskip

The classification of charge orbits, in particular the BPS ones, was
performed in \cite{FM} and \cite{DFL-0-brane}. By performing a suitable $%
U(1)\times SO\left( 6\right) \left( \sim U\left( 4\right) \right) $%
-transformation, the \textit{central charge matrix} $Z_{AB}$ can be \textit{%
skew-diagonalized} in the \textit{normal frame} (recall definition (\ref
{epsilon})):
\begin{equation}
Z_{AB}\overset{U(4)}{\longrightarrow }Z_{AB,skew-diag.}\equiv \left(
\begin{array}{cc}
z_{1} &  \\
& z_{2}
\end{array}
\right) \otimes \epsilon ,~~z_{1},z_{2}\in \mathbb{R}^{+},
\label{N=4-ZAB-normal-frame}
\end{equation}
where the ordering $z_{1}\geqslant z_{2}$ does not imply any loss of
generality. Furthermore, by performing a suitable $SO\left( M\right) $%
-transformation, the vector $Z_{I}$ of \textit{matter charges} can be
reduced to have only two non-vanishing entries, one real positive and the
other one complex, say (without loss of generality, with the subscript ``$%
red.$'' standing for ``\textit{reduced}'')
\begin{equation}
Z_{I}\overset{SO(M)}{\longrightarrow }Z_{I,red.}\equiv \left( \rho
_{1}e^{i\theta },\rho _{2},0,...,0\right) ,~~\rho _{1},\rho _{2}\in \mathbb{R%
}^{+},~~\theta \in \mathbb{R}.  \label{N=4-ZI-reduced}
\end{equation}
For non-vanishing (in general different) \textit{skew-eigenvalues} $z_{1}$
and $z_{2}$, the symmetry group of $Z_{AB,skew-diag.}$ is $\left( USp\left(
2\right) \right) ^{2}\sim \left( SU\left( 2\right) \right) ^{2}$.
Analogously, for non-vanishing (in general different) $\rho _{1}$ and $\rho
_{2}$ (and non-vanishing phase $\theta $) the symmetry group of $Z_{I,red.}$
is $SO\left( M-2\right) $. Thus, beside $z_{1}$, $z_{2}$, $\rho _{1}$, $\rho
_{2}$ and $\theta $ the generic $Z_{AB}$ and $Z_{I}$ are described by $%
7+2M=dim_{\mathbb{R}}\left( \frac{U\left( 4\right) \times SO\left( M\right)
}{\left( SU\left( 2\right) \right) ^{2}\times SO\left( M-2\right) }\right) $
\textit{``generalized angles''}. Consistently, the total number of
parameters is $2+2+1+7+2M=12+2M$, which is the real dimension of the
bi-fundamental representation $\left( \mathbf{2},\mathbf{6+M}\right) $,
defining the embedding of $SL\left( 2,\mathbb{R}\right) \times SO\left(
6,M\right) $ into $Sp\left( 12+2M,\mathbb{R}\right) $.\medskip

In $\mathcal{N}=4$, $d=4$ \textit{matter coupled} supergravity three
distinct \textit{large} charge orbits of the $\left( \mathbf{2},\mathbf{%
6+M}\right) $ of $SL\left( 2,\mathbb{R}\right) \times SO\left( 6,M\right) $
(for which $\mathcal{I}_{4,\mathcal{N}=4}\neq 0$, and the \textit{attractor
mechanism} holds) exist, as resulting from the analysis performed in%
\footnote{%
Consistent with the analysis of \cite{ADFT-review}, Eqs. (\ref
{N=4-1/4-BPS-large}), (\ref{N=4-non-BPS-Z=0-large}) and (\ref
{N=4-non-BPS-Z<>0-large}) fix a slightly misleading notation for the \textit{%
large} charge orbits of $\mathcal{N}=4$, $d=4$ \textit{matter} \textit{%
coupled} supergravity, as given by Table 1 of \cite{Kallosh-review}.} \cite
{ADFT-review}:

\begin{enumerate}
\item  The \textit{large }$\frac{1}{4}$-BPS orbit
\begin{equation}
\mathcal{O}_{\frac{1}{4}-BPS,\text{large}}=SL\left( 2,\mathbb{R}\right)
\times \frac{SO\left( 6,M\right) }{SO\left( 4,M\right) \times SO(2)},\text{~}%
dim_{\mathbb{R}}=11+2M,  \label{N=4-1/4-BPS-large}
\end{equation}
is defined by the $SL\left( 2,\mathbb{R}\right) \times SO\left( 6,M\right) $%
-invariant constraint
\begin{equation}
\mathcal{I}_{4,\mathcal{N}=4}>0.  \label{I4>0}
\end{equation}
Thus, the corresponding horizon solution of the $\mathcal{N}=4$, $d=4$
\textit{Attractor Eqs.} yields \cite{FM,DFL-0-brane,ADFT-review}
\begin{eqnarray}
z_{1} &\in &\mathbb{R}_{0}^{+},~~z_{2}=0,~~\rho _{1}=\rho _{2}=0,~~\theta ~%
\text{\textit{undetermined}};  \label{N=4-1/4-BPS-large-sol-1} \\
\mathcal{S}_{1} &=&z_{1}^{2}>0,~\mathcal{S}_{2}=0.
\label{N=4-1/4-BPS-large-sol-2}
\end{eqnarray}
Therefore, at the event horizon, the symmetry group of $Z_{AB,skew-diag.}$
defined in Eq. (\ref{N=4-ZAB-normal-frame}) does not get enhanced, while the
symmetry group of $Z_{i,red.}$ defined in Eq. (\ref{N=4-ZI-reduced}) gets
\textit{enhanced} as follows:
\begin{equation}
SO\left( M-2\right) \overset{r\rightarrow r_{H}^{+}}{\longrightarrow }%
SO\left( M\right) .  \label{N=4-1/4-BPS-large-enhancement}
\end{equation}
As a consequence, the horizon attractor solution exploits the maximal
compact symmetry $SU\left( 2\right) \times SU\left( 2\right) \times SO\left(
M\right) \times SO(2)$, which is the $mcs$ \cite{Gilmore} of the stabilizer
of $\mathcal{O}_{\frac{1}{4}-BPS,\text{large}}$ itself.

\item  The \textit{large }non-BPS $Z_{AB}=0$ orbit (existing for $%
M\geqslant 2$) \cite{ADFT-review}
\begin{equation}
\mathcal{O}_{non-BPS,Z_{AB}=0,\text{large}}=SL\left( 2,\mathbb{R}\right)
\times \frac{SO\left( 6,M\right) }{SO\left( 6,M-2\right) \times SO(2)},\text{%
~}dim_{\mathbb{R}}=11+2M,  \label{N=4-non-BPS-Z=0-large}
\end{equation}
is defined by the $SL\left( 2,\mathbb{R}\right) \times SO\left( 6,M\right) $%
-invariant constraint
\begin{equation}
\mathcal{I}_{4,\mathcal{N}=4}>0.
\end{equation}
Thus, the corresponding attractor solution of the $\mathcal{N}=4$, $d=4$
\textit{Attractor Eqs.} yields (for $M\geqslant 2$) \cite
{FM,DFL-0-brane,ADFT-review}
\begin{eqnarray}
z_{1} &=&z_{2}=0,~~\rho _{1}^{2}e^{2i\theta }+\rho _{2}^{2}=0\Leftrightarrow
\rho _{1}=\rho _{2}\in \mathbb{R}_{0}^{+},~\theta =\frac{\pi }{2}+k\pi
,~k\in \mathbb{Z};  \label{N=4-non-BPS-Z=0-large-sol-1} \\
\mathcal{S}_{1} &=&-2\rho _{1}^{2}<0,~\mathcal{S}_{2}=0.
\label{N=4-non-BPS-Z=0-large-sol-2}
\end{eqnarray}
Therefore, at the event horizon, the symmetry group of $Z_{AB,skew-diag.}$
defined in Eq. (\ref{N=4-ZAB-normal-frame}) gets enhanced as follows:
\begin{equation}
\left( SU\left( 2\right) \right) ^{2}\overset{r\rightarrow r_{H}^{+}}{%
\longrightarrow }SU\left( 4\right) ,
\label{N=4-non-BPS-ZAB=0-large-enhancement}
\end{equation}
and the symmetry group of $Z_{i,red.}$ defined in Eq. (\ref{N=4-ZI-reduced})
does not get \textit{enhanced}. Consequently, the horizon attractor solution
exploits the maximal compact symmetry $SU\left( 4\right) \times SO\left(
M-2\right) \times SO(2)$, which is the $mcs$ \cite{Gilmore} of the
stabilizer of $\mathcal{O}_{non-BPS,Z_{AB}=0,\text{large}}$ itself.

\item  The \textit{large }non-BPS $Z_{AB}\neq 0$ orbit (existing for $%
M\geqslant 1$) \cite{ADFT-review}
\begin{equation}
\mathcal{O}_{non-BPS,Z_{AB}\neq 0,\text{large}}=SL\left( 2,\mathbb{R}\right)
\times \frac{SO\left( 6,M\right) }{SO\left( 5,M-1\right) \times SO(1,1)},%
\text{~}dim_{\mathbb{R}}=11+2M,  \label{N=4-non-BPS-Z<>0-large}
\end{equation}
is defined by the $SL\left( 2,\mathbb{R}\right) \times SO\left( 6,M\right) $%
-invariant constraint
\begin{equation}
\mathcal{I}_{4,\mathcal{N}=4}<0.
\end{equation}
At the event horizon of the extremal BH, the solution of the $\mathcal{N}=4$%
, $d=4$ \textit{Attractor Eqs.} yields (for $M\geqslant 1$) \cite
{FM,DFL-0-brane,ADFT-review}
\begin{eqnarray}
z_{1} &=&z_{2}=\frac{\rho _{1}}{\sqrt{2}}\in \mathbb{R}_{0}^{+},~~\rho
_{2}=0,~~\theta =\frac{\pi }{2}+k\pi ,~k\in \mathbb{Z};
\label{N=4-non-BPS-Z<>0-large-sol-1} \\
\mathcal{S}_{1} &=&0,~\mathcal{S}_{2}=3z_{1}^{2}>0.
\label{N=4-non-BPS-Z<>0-large-sol-2}
\end{eqnarray}
Thus, at the event horizon, the symmetry group of $Z_{AB,skew-diag.}$
defined in Eq. (\ref{N=4-ZAB-normal-frame}) gets enhanced as follows:
\begin{equation}
\left( SU\left( 2\right) \right) ^{2}\overset{r\rightarrow r_{H}^{+}}{%
\longrightarrow }USp\left( 4\right) ,
\label{N=4-non-BPS-ZAB<>0-enhancement-1}
\end{equation}
and the symmetry group of $Z_{i,red.}$ defined in Eq. (\ref{N=4-ZI-reduced})
gets also \textit{enhanced} as
\begin{equation}
SO\left( M-2\right) \overset{r\rightarrow r_{H}^{+}}{\longrightarrow }%
SO\left( M-1\right) .  \label{N=4-non-BPS-ZAB<>0-enhancement-2}
\end{equation}
As a consequence, the horizon attractor solution exploits the maximal
compact symmetry $USp\left( 4\right) \times SO\left( M-1\right) $ which, due
to the isomorphism $USp\left( 4\right) \sim SO\left( 5\right) $, is the $mcs$
\cite{Gilmore} of the stabilizer of $\mathcal{O}_{non-BPS,Z_{AB}\neq 0,\text{%
large}}$ itself.\textbf{\medskip }
\end{enumerate}

As mentioned above, for such \textit{large }charge orbits, corresponding
to a non-vanishing quartic $SL\left( 2,\mathbb{R}\right) \times SO\left(
6,M\right) $-invariant $\mathcal{I}_{4,\mathcal{N}=4}$ and thus supporting
\textit{large} BHs, the \textit{attractor mechanism} holds.
Consequently, the computations of the Bekenstein-Hawking BH entropy can be
performed by solving the criticality conditions for the \textit{``BH
potential''}
\begin{equation}
V_{BH,\mathcal{N}=4}=\frac{1}{2}Z_{AB}\overline{Z}^{AB}+Z_{I}\overline{Z}%
^{I},  \label{N=4-VBH}
\end{equation}
the result being
\begin{equation}
\frac{S_{BH,}}{\pi }=\left. V_{BH,\mathcal{N}=4}\right| _{\partial V_{BH,%
\mathcal{N}=4}=0}=V_{BH,\mathcal{N}=4}\left( \phi _{H}\left( \mathcal{P}%
\right) ,\mathcal{P}\right) =\left| \mathcal{I}_{4,\mathcal{N}=4}\right|
^{1/2},  \label{N=4-AM}
\end{equation}
where $\phi _{H}\left( \mathcal{P}\right) $ denotes the set of solutions to
the \textit{criticality conditions} of $V_{BH,\mathcal{N}=4}$, namely the
\textit{Attractor Eqs.} of $\mathcal{N}=4$, $d=4$ \textit{matter coupled}
supergravity:
\begin{equation}
\partial _{\phi }V_{BH,\mathcal{N}=4}=0,~~\forall \phi \in \frac{SL\left( 2,%
\mathbb{R}\right) }{U(1)}\times \frac{SO\left( 6,M\right) }{SO\left(
6\right) \times SO\left( M\right) },  \label{N=4-AEs}
\end{equation}
expressing the stabilization of the scalar fields purely in terms of
supporting charges $\mathcal{P}$ at the event horizon of the extremal BH.
Through Eqs. (\ref{N=4-MC-1})-(\ref{N=4-MC-2}) and (\ref{N=4-VBH}), Eqs. (%
\ref{N=4-AEs}) can be rewritten as follows \cite{ADFT-review}:
\begin{equation}
\left\{
\begin{array}{l}
\left( \overline{Z}^{AB}+\frac{1}{2}\epsilon ^{ABCD}Z_{CD}\right) Z^{I}=0;
\\
\\
Z^{I}Z^{J}\delta _{IJ}+\frac{1}{4}\epsilon _{ABCD}\overline{Z}^{AB}\overline{%
Z}^{CD}=0.
\end{array}
\right.  \label{N=4-AEs-2}
\end{equation}

Actually, the critical potential $\left. V_{BH,\mathcal{N}=4}\right|
_{\partial V_{BH,\mathcal{N}=4}=0}$ exhibits some \textit{``flat''}
directions, so not all scalars are stabilized in terms of charges at the
event horizon \cite{Kallosh-review}. Thus, Eq. (\ref{N=4-AM}) yields that
the \textit{unstabilized} scalars, spanning a related \textit{moduli space}
of the considered class of attractor solutions, do not enter in the
expression of the BH entropy at all. The \textit{moduli spaces} exhibited by
the \textit{Attractor Eqs.} (\ref{N=4-AEs})-(\ref{N=4-AEs-2}) are \cite
{Kallosh-review}
\begin{eqnarray}
\mathcal{M}_{\frac{1}{4}-BPS,\text{large}} &=&\frac{SO\left( 4,M\right) }{%
SU\left( 2\right) \times SU\left( 2\right) \times SO\left( M\right) },~dim_{%
\mathbb{R}}=4M; \\
\mathcal{M}_{non-BPS,Z_{AB}=0,\text{large}} &=&\frac{SO\left( 6,M-2\right) }{%
SU\left( 4\right) \times SO\left( M-2\right) },~dim_{\mathbb{R}}=6\left(
M-2\right) ; \\
\mathcal{M}_{non-BPS,Z_{AB}\neq 0,\text{large}} &=&SO\left( 1,1\right)
\times \frac{SO\left( 5,M-1\right) }{USp\left( 4\right) \times SO\left(
M-1\right) },~dim_{\mathbb{R}}=5\left( M-1\right) +1.  \notag \\
&&  \label{Stanford-2}
\end{eqnarray}
As justified in \cite{ADF-U-duality-d=4} and then in \cite{Kallosh-review}, $%
\mathcal{M}_{\frac{1}{4}-BPS,\text{large}}$ is a \textit{quaternionic}
symmetric manifold. Furthermore, $\mathcal{M}_{non-BPS,Z_{AB}\neq 0,\text{%
large}}$ given by Eq. (\ref{Stanford-2}) is nothing but the scalar manifold
of $\mathcal{N}=4$, $d=5$ \textit{matter coupled} supergravity. The
stabilizers of $\mathcal{M}_{\frac{1}{4}-BPS,\text{large}}$, $\mathcal{M}%
_{non-BPS,Z_{AB}=0,\text{large}}$ and $\mathcal{M}_{non-BPS,Z_{AB}\neq 0,%
\text{large}}$ exploit the maximal compact symmetry of the corresponding
charge orbits; this symmetry becomes fully manifest through the enhancement
of the compact symmetry group of $Z_{AB,skew-diag.}$ and $Z_{I,red.}$ at the
event horizon of the extremal BH, respectively given by Eqs. (\ref
{N=4-1/4-BPS-large-enhancement}), (\ref{N=4-non-BPS-ZAB=0-large-enhancement}%
) and (\ref{N=4-non-BPS-ZAB<>0-enhancement-1})-(\ref
{N=4-non-BPS-ZAB<>0-enhancement-2}).\bigskip

Let us now analyze the \textit{small} charge orbits of the $\left(
\mathbf{2},\mathbf{6+M}\right) $ of $SL\left( 2,\mathbb{R}\right) \times
SO\left( 6,M\right) $, associated to $\mathcal{I}_{4,\mathcal{N}=4}=0$, for
which the \textit{attractor mechanism} does not hold. The analysis performed
below completes the one given in \cite{FM} and \cite{DFL-0-brane}.

While in $\mathcal{N}=8$, $d=4$ supergravity all three \textit{small}
charge orbits are BPS (with various degrees of supersymmetry-preservation), in the considered $\mathcal{N}=4$, $d=4$ theory there are five
\textit{small} charge orbits, two of them being $\frac{1}{2}$-BPS, 
one $\frac{1}{4}$-BPS, and the other two non-BPS (one with $Z_{AB}=0$ 
and the other with $Z_{AB}\neq 0$). Such an abundance of different charge orbits can be
traced back to the \textit{factorized} nature of the $U$-duality group $%
SL\left( 2,\mathbb{R}\right) \times SO\left( 6,M\right) $. Furthermore, it
should be remarked that in $\mathcal{N}=4$, $d=4$ supergravity the $\frac{1}{%
\left( \mathcal{N}=\right) 4}$-BPS charge orbit exists only in its \textit{%
large} version, differently from the $d=4$ maximal theory, in which both
\textit{large} and \textit{small} $\frac{1}{\left( \mathcal{N}%
=\right) 8}$-BPS charge orbits exist.

It is now convenient to denote with $\alpha _{1}$ and $\alpha _{2}$ the two
real non-negative eigenvalues of the matrix $Z_{AB}\overline{Z}^{CB}=\left(
ZZ^{\dag }\right) _{A}^{C}$. By recalling Eq. (\ref{N=4-ZAB-normal-frame}),
one can notice that ($i=1,2$)
\begin{equation}
\alpha _{i}=z_{i}^{2}.
\end{equation}
and one can order them as $\alpha _{1}\geqslant \alpha _{2}$, without any
loss of generality.\smallskip\ The explicit expression of $\alpha _{i}$ in
terms of $U\left( 4\right) \times SO\left( M\right) $-invariants (namely of $%
Tr\left( ZZ^{\dag }\right) $, $Tr\left( \left( ZZ^{\dag }\right) ^{2}\right)
$, and suitable powers) is given by Eqs. (5.108) and (5.109) of \cite
{DFL-0-brane}.

Firstly, let us observe that from Eqs. (\ref{N=4-bare}) and (\ref
{N=4-dressed}) the $SL\left( 2,\mathbb{R}\right) \times SO\left( 6,M\right) $%
-invariant \textit{``degeneracy''} condition can be written in the \textit{%
``dressed''} ($\mathcal{R}$-symmetry- and $SO\left( M\right) $- covariant)
and \textit{``bare''} (symplectic-, \textit{i.e.} $Sp\left( 12+2M,\mathbb{R}%
\right) $- covariant) charges' bases respectively as follows:
\begin{equation}
\mathcal{I}_{4,\mathcal{N}=4}=0\Leftrightarrow \mathcal{S}_{1}^{2}=\left|
\mathcal{S}_{2}\right| ^{2}\Leftrightarrow p^{2}q^{2}=\left( p\cdot q\right)
^{2}\geqslant 0.  \label{N=4-small}
\end{equation}

Then, in order to determine the number and typology of \textit{small}
orbits, it is convenient to start differentiating $\mathcal{I}_{4,\mathcal{N}%
=4}$ in the symplectic \textit{``bare''} charges' basis $\mathcal{P}\equiv
\left( p^{\Lambda },q_{\Lambda }\right) ^{T}$ (recall definition (\ref{P})).
Eqs. (\ref{N=4-bare}) and (\ref{T-tensor-antisymm}) yield the constraints
defining the ``\textit{small'' critical} orbits to read
\begin{eqnarray}
\frac{\partial \mathcal{I}_{4,\mathcal{N}=4}}{\partial p_{\Lambda }} &=&2%
\left[ q^{2}p^{\Lambda }-\left( q\cdot p\right) q^{\Lambda }\right]
=2T^{\left( a\right) \mid \Lambda \Sigma }q_{\Sigma }=0;  \label{N=4-crit-1}
\\
\frac{\partial \mathcal{I}_{4,\mathcal{N}=4}}{\partial q_{\Lambda }} &=&2%
\left[ p^{2}q^{\Lambda }-\left( q\cdot p\right) p^{\Lambda }\right]
=-2T^{\left( a\right) \mid \Lambda \Sigma }p_{\Sigma }=0.  \label{N=4-crit-2}
\end{eqnarray}
Due to the definition (\ref{T-tensor-antisymm}), or equivalently to the
homogeneity (of degree four) in charges of\textbf{\ }$\mathcal{I}_{4,%
\mathcal{N}=4}$,\ it is worth noticing that the \textit{``criticality''}
constraints (\ref{N=4-crit-1}) and (\ref{N=4-crit-2}) imply the \textit{%
``degeneracy''} condition (\ref{N=4-small}).

Beside the trivial one ($p_{\Lambda }=0=q_{\Lambda }~\forall \Lambda $), all
the solutions to the \textit{``criticality''} constraints (\ref{N=4-crit-1})
and (\ref{N=4-crit-2}) list as follows:
\begin{eqnarray}
&&A]~~\left\{
\begin{array}{l}
T_{\Lambda \Sigma }^{\left( a\right) }=0; \\
\\
\left\{
\begin{array}{l}
p^{2}q^{2}=\left( p\cdot q\right) ^{2}>0:~\left\{
\begin{array}{l}
A.1]~~p^{2}>0,~q^{2}>0; \\
\mathit{aut} \\
A.2]~~p^{2}<0,~q^{2}<0;
\end{array}
\right. \\
\\
~A.3]~~p^{2}q^{2}=\left( p\cdot q\right) ^{2}=0:~~p^{2}=0,~q^{2}=0;
\end{array}
\right.
\end{array}
\right. ~  \label{A} \\
&&  \notag \\
&&B]~~\left\{
\begin{array}{l}
T_{\Lambda \Sigma }^{\left( a\right) }\neq 0; \\
\\
p^{2}=q^{2}=p\cdot q=0.
\end{array}
\right. ~  \label{B}
\end{eqnarray}
Notice that each set (\textbf{A.1}, \textbf{A.2}, \textbf{A.3} and \textbf{B}%
) of constraints is $SL\left( 2,\mathbb{R}\right) \times SO\left( 6,M\right)
$-invariant, but formulated in terms of the symplectic charge basis $%
\mathcal{P}$.

The solutions (\ref{A})-(\ref{B}) can be rewritten by noticing that $\frac{%
\partial ^{2}\mathcal{I}_{4,\mathcal{N}=4}}{\partial \mathcal{P}\partial
\mathcal{P}}$, \textit{i.e.} the tensor of second derivatives of $\mathcal{I}%
_{4,\mathcal{N}=4}$ with respect to $\mathcal{P},$ sits in the symmetric
product representation $\left( \left( \mathbf{2},\mathbf{6+M}\right) \times
\left( \mathbf{2},\mathbf{6+M}\right) \right) _{s}$ of the $U$-duality group
$SL\left( 2,\mathbb{R}\right) \times SO\left( 6,M\right) $, which decomposes
as follows \cite{DFL-0-brane}:
\begin{equation}
\left( \left( \mathbf{2},\mathbf{6+M}\right) \times \left( \mathbf{2},%
\mathbf{6+M}\right) \right) _{s}\overset{SL\left( 2,\mathbb{R}\right) \times
SO\left( 6,M\right) }{\longrightarrow }\underset{T^{\left( 0\right) }}{%
\left( \mathbf{3},\mathbf{1}\right) }+\underset{T_{\Lambda \Sigma }^{\left(
tr-s\right) }}{\left( \mathbf{3},\mathbf{TrSym}\left( SO(6,M)\right) \right)
}+\underset{T_{\Lambda \Sigma }^{\left( a\right) }}{\left( \mathbf{1},%
\mathbf{Adj}\left( SO(6,M)\right) \right) }.  \label{N=4-decomp}
\end{equation}
The antisymmetric tensor
\begin{equation}
T_{\Lambda \Sigma }^{\left( a\right) }\equiv \left. \frac{\partial ^{2}%
\mathcal{I}_{4,\mathcal{N}=4}}{\partial \mathcal{P}\partial \mathcal{P}}%
\right| _{\left( \mathbf{1},\mathbf{Adj}\left( SO(6,M)\right) \right) }
\label{T-tensor-antisymm-2}
\end{equation}
was already introduced in Eq. (\ref{T-tensor-antisymm}). $\mathbf{TrSym}$
and $\mathbf{Adj}$ respectively denote the \textit{traceless symmetric} and
\textit{adjoint} representations, and\textbf{\ }\cite{DFL-0-brane}
\begin{eqnarray}
T_{\Lambda \Sigma }^{\left( tr-s\right) } &\equiv &\left. \frac{\partial ^{2}%
\mathcal{I}_{4,\mathcal{N}=4}}{\partial \mathcal{P}\partial \mathcal{P}}%
\right| _{\left( \mathbf{3},\mathbf{TrSym}\left( SO(6,M)\right) \right)
}\equiv  \notag \\
&\equiv &\left( q_{\Lambda }q_{\Sigma }-\frac{q^{2}}{6+M}\eta _{\Lambda
\Sigma },~p_{\Lambda }p_{\Sigma }-\frac{p^{2}}{6+M}\eta _{\Lambda \Sigma },~%
\frac{1}{2}\left( q_{\Lambda }p_{\Sigma }+q_{\Sigma }p_{\Lambda }\right) -%
\frac{q\cdot p}{6+M}\eta _{\Lambda \Sigma }\right) ;  \notag \\
&&  \label{T-tr-s} \\
&&  \notag \\
T^{\left( 0\right) } &\equiv &\left. \frac{\partial ^{2}\mathcal{I}_{4,%
\mathcal{N}=4}}{\partial \mathcal{P}\partial \mathcal{P}}\right| _{\left(
\mathbf{3},\mathbf{1}\right) }\equiv Tr_{SO\left( 6,M\right) }\left(
T_{\Lambda \Sigma }^{\left( s\right) }\right) \equiv Tr_{SO\left( 6,M\right)
}\left( \left. \frac{\partial ^{2}\mathcal{I}_{4,\mathcal{N}=4}}{\partial
\mathcal{P}\partial \mathcal{P}}\right| _{\left( \mathbf{3},\mathbf{Sym}%
\left( SO(6,M)\right) \right) }\right) =  \notag \\
&=&\left( q^{2},p^{2},q\cdot p\right) =\left(
\begin{array}{ccc}
q^{2} &  & q\cdot p \\
&  &  \\
q\cdot p &  & p^{2}
\end{array}
\right) .  \label{T0}
\end{eqnarray}
The definition (\ref{T0}) of $T^{\left( 0\right) }$ implies that (recall Eq.
(\ref{N=4-bare}))
\begin{equation}
\mathcal{I}_{4,\mathcal{N}=4}=det\left( T^{\left( 0\right) }\right)
=det\left( \left. \frac{\partial ^{2}\mathcal{I}_{4,\mathcal{N}=4}}{\partial
\mathcal{P}\partial \mathcal{P}}\right| _{\left( \mathbf{3},\mathbf{1}%
\right) }\right) ,  \label{N=4-G-inv-rewritten}
\end{equation}
in turn yielding another, equivalent $SL\left( 2,\mathbb{R}\right) \times
SO\left( 6,M\right) $-invariant characterization of the \textit{%
``degeneracy''} condition (\ref{N=4-small}):
\begin{equation}
det\left( T^{\left( 0\right) }\right) =det\left( \left. \frac{\partial ^{2}%
\mathcal{I}_{4,\mathcal{N}=4}}{\partial \mathcal{P}\partial \mathcal{P}}%
\right| _{\left( \mathbf{3},\mathbf{1}\right) }\right) =0.
\label{N=4-small-equivalent}
\end{equation}
Thus, Eqs. (\ref{A})-(\ref{B}) can be recast as follows:
\begin{eqnarray}
&&A]~~\left\{
\begin{array}{l}
T_{\Lambda \Sigma }^{\left( a\right) }=0; \\
\\
det\left( T^{\left( 0\right) }\right) =0,~\left\{
\begin{array}{l}
A.1]~~Tr\left( T^{\left( 0\right) }\right) >0; \\
\mathit{aut} \\
A.2]~~Tr\left( T^{\left( 0\right) }\right) <0; \\
\mathit{aut} \\
A.3]~Tr\left( T^{\left( 0\right) }\right) =0\Leftrightarrow T^{\left(0\right) }=0;
\end{array}
\right.
\end{array}
\right. ~~  \label{A-1} \\
&&  \notag \\
&&B]~~\left\{
\begin{array}{l}
T_{\Lambda \Sigma }^{\left( a\right) }\neq 0; \\
\\
T^{\left(0\right) }=0.
\end{array}
\right. ~  \label{B-1}
\end{eqnarray}
As mentioned above, each set (\textbf{A.1}, \textbf{A.2}, \textbf{A.3} and
\textbf{B}) of constraints is $SL\left( 2,\mathbb{R}\right) \times SO\left(
6,M\right) $-invariant, but formulated in terms of the symplectic charge
basis $\mathcal{P}$.

It is interesting to point out that, differently from $\mathcal{N}=8$, $d=4$
supergravity treated in Sect. \ref{N=8-ungauged}, in $\mathcal{N}=4$, $d=4$
supergravity there are no \textit{small doubly-critical }(or with higher
degree of criticality) charge orbits \textit{independent} from the \textit{%
small critical} ones. This can be easily seen by noticing that the
solutions (\ref{A-1})-(\ref{B-1}) to the \textit{``criticality''}
constraints (\ref{N=4-crit-1}) and (\ref{N=4-crit-2}) can actually be
rewritten in a \textit{doubly-critical} fashion, \textit{i.e.} through $%
\frac{\partial ^{2}\mathcal{I}_{4,\mathcal{N}=4}}{\partial \mathcal{P}%
\partial \mathcal{P}}$ and related projections (according to decomposition (%
\ref{N=4-decomp})). For completeness' sake, we report here the second order
derivatives of $\mathcal{I}_{4,\mathcal{N}=4}$ with respect to the \textit{%
``bare'' symplectic }charges:
\begin{eqnarray}
\frac{\partial ^{2}\mathcal{I}_{4,\mathcal{N}=4}}{\partial p_{\Sigma
}\partial p_{\Lambda }} &=&2\left( q^{2}\eta ^{\Lambda \Sigma }-q^{\Lambda
}q^{\Sigma }\right) ; \\
\frac{\partial ^{2}\mathcal{I}_{4,\mathcal{N}=4}}{\partial q_{\Sigma
}\partial q_{\Lambda }} &=&2\left( p^{2}\eta ^{\Lambda \Sigma }-p^{\Lambda
}p^{\Sigma }\right) ; \\
\frac{\partial ^{2}\mathcal{I}_{4,\mathcal{N}=4}}{\partial q_{\Sigma
}\partial p_{\Lambda }} &=&4T^{\left( a\right) \mid \Lambda \Sigma }.
\end{eqnarray}
\medskip

In order to determine the \textit{small }orbits of the bi-fundamental
representation $(\mathbf{2},\mathbf{6+M})$ of the $U$-duality group $%
SL\left( 2,\mathbb{R}\right) \times SO\left( 6,M\right) $ and to study their
supersymmetry-preserving properties, it is now convenient to switch to the
basis of \textit{``dressed''} charges (recall Eqs. (\ref{Z-1}) and (\ref{Z-2}%
))
\begin{equation}
\mathcal{U}\equiv \left( \mathcal{Z},\overline{\mathcal{Z}}\right)
^{T}=\left( Z_{AB,}Z^{I},\overline{Z}_{AB},\overline{Z}^{I}\right) ^{T}.
\end{equation}
From the analysis of \cite{DFL-0-brane}, one obtains the following
equivalence:
\begin{equation}
T_{\Lambda \Sigma }^{\left( a\right) }\equiv \left. \frac{\partial ^{2}%
\mathcal{I}_{4,\mathcal{N}=4}}{\partial \mathcal{P}\partial \mathcal{P}}%
\right| _{\mathbf{Adj}\left( SO(6,M)\right) }=0\Leftrightarrow \left. \frac{%
\partial ^{2}\mathcal{I}_{4,\mathcal{N}=4}}{\partial \mathcal{U}\partial
\mathcal{U}}\right| _{\mathbf{Adj}\left( SO(6,M)\right) }=0.
\label{T-symm=0}
\end{equation}
The $SL\left( 2,\mathbb{R}\right) \times SO\left( 6,M\right) $-invariant
constraint (\ref{T-symm=0}) is common to the \textit{small critical}
charge orbits determined by the solutions \textbf{A.1}, \textbf{A.2} and
\textbf{A.3} of Eqs. (\ref{A-1}). It also implies that $\alpha _{1}=\alpha
_{2}$ \cite{DFL-0-brane}. Then, the further $SL\left( 2,\mathbb{R}\right)
\times SO\left( 6,M\right) $-invariant constraints $Tr\left( T^{\left(
0\right) }\right) \gtreqless 0$ can equivalently be rewritten as (recall
definition (\ref{S1}))
\begin{equation}
Tr\left( T^{\left( 0\right) }\right) \gtreqless 0\Leftrightarrow \mathcal{S}%
_{1}\gtreqless 0.
\end{equation}
Therefore, one can characterize the \textit{small~critical}$\mathit{~}$%
orbits \textbf{A.1}, \textbf{A.2} and \textbf{A.3} of Eqs. (\ref{A}) and (%
\ref{A-1}) as follows:
\begin{equation}
A]~~\left\{
\begin{array}{l}
\left. \frac{\partial ^{2}\mathcal{I}_{4,\mathcal{N}=4}}{\partial \mathcal{U}%
\partial \mathcal{U}}\right| _{\mathbf{Adj}\left( SO(6,M)\right) }=0; \\
\\
\mathcal{S}_{1}^{2}=\left| \mathcal{S}_{2}\right| ^{2},~\left\{
\begin{array}{l}
A.1]~~\mathcal{S}_{1}>0; \\
\mathit{aut} \\
A.2]~~\mathcal{S}_{1}<0; \\
\mathit{aut} \\
A.3]~\mathcal{S}_{1}=0\Leftrightarrow \mathcal{S}_{2}=0. 
\end{array}
\right.
\end{array}
\right.  \label{A-2}
\end{equation}
Notice that each set (\textbf{A.1}, \textbf{A.2}, \textbf{A.3} and \textbf{B}%
) of constraints is $SL\left( 2,\mathbb{R}\right) \times SO\left( 6,M\right)
$-invariant but, differently from Eqs. (\ref{A}) and (\ref{A-1}), it is also
independent from the symplectic basis eventually considered.

On the other hand, the $SL\left( 2,\mathbb{R}\right) \times SO\left(
6,M\right) $-invariant constraints (\ref{B}) and (\ref{B-1}) defining the
\textit{small~critical}$\mathit{~}$orbit \textbf{B} can be recast in a
form which (differently from Eqs. (\ref{B}) and (\ref{B-1})) is independent
from the symplectic basis eventually considered, as follows:
\begin{equation}
B]~~\left\{
\begin{array}{l}
\left. \frac{\partial ^{2}\mathcal{I}_{4,\mathcal{N}=4}}{\partial \mathcal{U}%
\partial \mathcal{U}}\right| _{\mathbf{Adj}\left( SO(6,M)\right) }\neq 0; \\
\\
\mathcal{S}_{1}^{2}=\left| \mathcal{S}_{2}\right| ^{2}=0.
\end{array}
\right. ~  \label{B-2}
\end{equation}

Thus, five distinct \textit{small} charge orbits (all with $\mathcal{I}%
_{4,\mathcal{N}=4}=0$) exist:

\begin{enumerate}
\item  The \textit{critical} orbit \textbf{A.1} is defined by the $SL\left(
2,\mathbb{R}\right) \times SO\left( 6,M\right) $-invariant constraints (\ref
{A}) (or (\ref{A-1}), or (\ref{A-2})). Such constraints are solved by the
following flow solution (exhibiting maximal symmetry):
\begin{equation}
z_{1}=z_{2}\in \mathbb{R}_{0}^{+},~\rho _{1}=\rho _{2}=0,~\theta ~\text{%
\textit{undetermined}.}  \label{N=4-1/2-BPS-sol}
\end{equation}
Thus, from the reasoning performed at the end of Sect. \ref{Duality} and the
analysis of \cite{DFL-0-brane}, the considered \textit{small critical}
orbit is $\frac{1}{2}$-BPS. Along the corresponding \textit{small
critical }$\frac{1}{2}$-BPS flow, the (maximal compact) symmetry of the
\textit{skew-diagonalized} central charge matrix $Z_{AB,skew-diag.}$ defined
in Eq. (\ref{N=4-ZAB-normal-frame}) is $USp\left( 4\right) $, whereas the
one of $Z_{I,red.}$ defined in Eq. (\ref{N=4-ZI-reduced}) is $SO\left(
M\right) $. Therefore, the resulting maximal compact symmetry of the \textit{%
critical} orbit \textbf{A.1 }is $USp\left( 4\right) \times SO\left( M\right)
$.

\item  The \textit{critical} orbit \textbf{A.2} is defined by the $SL\left(
2,\mathbb{R}\right) \times SO\left( 6,M\right) $-invariant constraints (\ref
{A}) (or (\ref{A-1}), or (\ref{A-2})). Such constraints are solved by the
following flow solution, existing for $M\geqslant 1$ (and exhibiting maximal
symmetry)
\begin{equation}
z_{1}=z_{2}=0,~\rho _{1}\in \mathbb{R}_{0}^{+},~\rho _{2}=0.
\label{N=4-non-BPS-Z=0-small-sol-1}
\end{equation}
Thus, the considered \textit{small critical} orbit is non-BPS $Z_{AB}=0$%
. Along the corresponding \textit{small critical }non-BPS $Z_{AB}=0$
flow, the (maximal compact) symmetry of the \textit{skew-diagonalized}
central charge matrix $Z_{AB,skew-diag.}$ defined in Eq. (\ref
{N=4-ZAB-normal-frame}) is $SU\left( 4\right) $, whereas the one of $%
Z_{I,red.}$ defined in Eq. (\ref{N=4-ZI-reduced}) is $SO\left( M-1\right) $.
Therefore, the resulting maximal compact symmetry of the \textit{critical}
orbit \textbf{A.2 }is $SU\left( 4\right) \times SO\left( M-1\right) $.

\item  The \textit{critical} orbit \textbf{A.3} is defined by the $SL\left(
2,\mathbb{R}\right) \times SO\left( 6,M\right) $-invariant constraints (\ref
{A}) (or (\ref{A-1}), or (\ref{A-2})). Such constraints are solved by the
following flow solution, existing for\textbf{\ }$M\geqslant 1$\textbf{\ (}%
and exhibiting maximal symmetry)
\begin{equation}
z_{1}=z_{2}=\frac{\rho _{2}}{\sqrt{2}}\in \mathbb{R}_{0}^{+},~\rho
_{1}=0,~\theta ~\text{\textit{undetermined}.}
\label{N=4-non-BPS-Z<>0-small-I-sol-1}
\end{equation}
This \textit{small critical} orbit is $\frac{1}{2}$-BPS.
Along the corresponding \textit{small critical }non-BPS $%
Z_{AB}\neq 0$ flow, the (maximal compact) symmetry of the \textit{%
skew-diagonalized} central charge matrix $Z_{AB,skew-diag.}$ defined in Eq. (%
\ref{N=4-ZAB-normal-frame}) is $USp\left( 4\right) $, whereas the one of $%
Z_{I,red.}$ defined in Eq. (\ref{N=4-ZI-reduced}) is $SO\left( M-1\right) $.
Therefore, the resulting maximal compact symmetry of the \textit{critical}
orbit \textbf{A.3 }is $USp\left( 4\right) \times SO\left( M-1\right) $.

\item  The \textit{critical} orbit \textbf{B} is defined by the $SL\left( 2,%
\mathbb{R}\right) \times SO\left( 6,M\right) $-invariant constraints (\ref{B}%
) (or (\ref{B-1}), or (\ref{B-2})). Such constraints are solved by the
following flow solution, existing for $M\geqslant 2$ (and exhibiting maximal
symmetry)
\begin{eqnarray}
z_{1} &\in &\mathbb{R}_{0}^{+},~z_{2}=0,~\rho _{1}=\rho _{2}=\frac{z_{1}}{%
\sqrt{2}};  \label{N=4-non-BPS-Z<>0-small-II-sol-1} \\
~\theta &=&\frac{\pi }{2}+k\pi ,~k\in \mathbb{Z}.
\label{N=4-non-BPS-Z<>0-small-II-sol-2}
\end{eqnarray}
This \textit{small critical} orbit is $\frac{1}{4}$-BPS.
Along the corresponding \textit{small critical }non-BPS $%
Z_{AB}\neq 0$ flow, the (maximal compact) symmetry of the \textit{%
skew-diagonalized} central charge matrix $Z_{AB,skew-diag.}$ defined in Eq. (%
\ref{N=4-ZAB-normal-frame}) is $\left( SU\left( 2\right) \right) ^{2}$,
whereas the one of $Z_{I,red.}$ defined in Eq. (\ref{N=4-ZI-reduced}) is $%
SO\left( M-2\right) $. Therefore, the resulting maximal compact symmetry of
the \textit{critical} orbit \textbf{B }is $\left( SU\left( 2\right) \right)
^{2}\times SO\left( M-2\right) $.

\item  The generic \textit{small lightlike} case is defined by the $%
SL\left( 2,\mathbb{R}\right) \times SO\left( 6,M\right) $-invariant
constraints (\ref{N=4-small}) (or (\ref{N=4-small-equivalent})). In this
case, it is more convenient to consider the symplectic basis of \textit{%
``bare''} charges $\mathcal{P}$ and, in order to determine the maximal
compact symmetry of the flow solution(s), one can consider the saturation of
the bound (\ref{N=4-small}), namely:
\begin{equation}
p^{2}q^{2}=\left( p\cdot q\right) ^{2}=0.
\end{equation}
This is in general solved by $p^{2}=0$, $p\cdot q=0$ and $q^{2}\neq 0$ (or
equivalently by $q^{2}=0$, $p\cdot q=0$ and $p^{2}\neq 0$). It is easy to
realize that the maximal compact symmetry of the flow solution is $SO\left(
4\right) \times SO\left( M-1\right) $ in the case $q^{2}>0$, and $SO\left(
5\right) \times SO\left( M-2\right) $ in the case $q^{2}<0$. In the first
case the solution exists for $M\geqslant 1$, whereas in the second case the
solution exists for $M\geqslant 2$. Thus, one actually gets two generic
\textit{small lightlike} orbits, both non-BPS $Z_{AB}\neq 0$, with
maximal compact symmetry respectively given by $SO\left( 4\right) \times
SO\left( M-1\right) $ and $SO\left( 5\right) \times SO\left( M-2\right) $%
.\medskip
\end{enumerate}

\textit{Mutatis mutandis}, the same considerations made at the end of Sect.
\ref{N=8-ungauged} for $\mathcal{N}=8$, $d=4$ supergravity also hold for $%
\mathcal{N}=4$, $d=4$ \textit{matter coupled} supergravity.

Notice that in \textit{pure} $\mathcal{N}=4$, $d=4$ supergravity only the \textit{small} $\frac{1}{2}$-BPS orbit \textbf{A.1} and the \textit{large} $%
\frac{1}{4}$-BPS orbit exist. Indeed, the non-BPS $Z_{AB}\neq 0$ and non-BPS $Z_{AB}=0$ \textit{large} orbits and the \textit{small} orbits \textbf{A.2},
\textbf{A.3} and \textbf{B} cannot be realized, and the \textit{small lightlike} orbit(s) of point 5 above coincide with \textit{small} orbit \textbf{A.1}.

Finally, it is worth noticing that the $U\left( 1\right) $ (stabilizer of
the factor $\frac{SL\left( 2,\mathbb{R}\right) }{U\left( 1\right) }$ of the
scalar manifold (\ref{N=4-scalar-manifold})) is broken both in \textit{%
large} and \textit{small} charge orbits, because both the \textit{%
central charge matrix} $Z_{AB}$ and the \textit{matter charges} $Z_{I}$ are
charged with respect to it.

\section{\label{N=2-symmetric-ungauged}$\mathcal{N}=2$}

In $\mathcal{N}=2$, $d=4$ supergravity one can repeat the analysis of \cite
{GZ,AFZ} (see also \cite{de-Wit-review}), by using the properties of \textit{%
special K\"{a}hler geometry} (SKG, see \textit{e.g.} \cite{CDF-review}, and
Refs. therein). Indeed, in SKG one can define an $Sp\left( 2n,\mathbb{R}%
\right) $ matrix over the scalar manifold (as in Eq. (\ref{CERNN-1})), as
well complex matrices $f$ and $h$ (as in Eqs. (\ref{f-h-1})-(\ref{CERNN-2}%
)), without the need for the manifold to be necessarily a(n \textit{at least
locally}) symmetric space (see \textit{e.g.} \cite{ADF-central,ADFT-review}).

The basic identities of SKG applied to the (covariantly holomorphic) $%
\mathcal{N}=2$, $d=4$ \textit{central charge} section
\begin{equation}
Z\equiv e^{K/2}\left( X^{\Lambda }q_{\Lambda }-F_{\Lambda }p^{\Lambda
}\right)
\end{equation}
of the $U\left( 1\right) $ K\"{a}hler-Hodge bundle (with K\"{a}hler weights $%
\left( 1,-1\right) $) read as follows \cite{FGK} ($i,\overline{j}=1,...,n-1$%
, with $n-1$ denoting the number of Abelian vector multiplets coupled to the
supergravity one)
\begin{eqnarray}
\overline{D}_{\overline{i}}Z &=&0;  \label{SKG-id-1} \\
D_{i}D_{j}Z &=&iC_{ijk}g^{k\overline{k}}\overline{D}_{\overline{k}}\overline{%
Z};  \label{SKG-id-2} \\
\overline{D}_{\overline{j}}D_{i}Z &=&g_{i\overline{j}}Z,  \label{SKG-id-3}
\end{eqnarray}
where $\left( X^{\Lambda },F_{\Lambda }\right) $ are the holomorphic
symplectic sections of the $U\left( 1\right) $ K\"{a}hler-Hodge bundle (with
K\"{a}hler weights $\left( 2,0\right) $), and $K$ denotes the K\"{a}hler
potential of the Abelian vector multiplets' scalar manifold, with metric $%
g_{i\overline{j}}=\overline{\partial }_{\overline{j}}\partial _{i}K$. $%
C_{ijk}$ is the rank-$3$ symmetric and covariantly holomorphic $C$\textit{%
-tensor} of SKG (see \textit{e.g.} \cite{CDF-review}, and Refs. therein):
\begin{eqnarray}
\overline{D}_{\overline{l}}C_{ijk} &=&0;  \label{C-1} \\
D_{[l}C_{i]jk} &=&0.  \label{C-2}
\end{eqnarray}

Thus, in $\mathcal{N}=2$, $d=4$ supergravity coupled to $n-1$ Abelian vector
multiplets, the \textit{``BH potential"} is given by \cite{FK1,FK2}
\begin{equation}
V_{BH}\left( \phi ,\mathcal{P}\right) =Z\overline{Z}+g^{i\overline{j}}\left(
D_{i}Z\right) \overline{D}_{\overline{j}}\overline{Z},  \label{VBH-2}
\end{equation}
and the \textit{Attractor Eqs.} read \cite{FGK}
\begin{equation}
\partial _{i}V_{BH}=0\Leftrightarrow 2\overline{Z}D_{i}Z+iC_{ijk}g^{j%
\overline{j}}g^{k\overline{k}}\left( \overline{D}_{\overline{j}}\overline{Z}%
\right) \overline{D}_{\overline{k}}\overline{Z}=0.  \label{N=2-d=4-AEs}
\end{equation}

\begin{enumerate}
\item  The ($\frac{1}{2}$-BPS) supersymmetric solution to \textit{Attractor
Eqs. }(\ref{N=2-d=4-AEs}) is determined by
\begin{equation}
\left( D_{i}Z\right) _{\frac{1}{2}-BPS}=0,~\forall i,  \label{1/2-BPS-1}
\end{equation}
and therefore Eq. (\ref{VBH-2}) yields
\begin{equation}
V_{BH,\frac{1}{2}-BPS}=\left| Z\right| _{\frac{1}{2}-BPS}^{2},
\label{1/2-BPS-2}
\end{equation}
and the corresponding Hessian matrix of $V_{BH}$ has block components given
by \cite{FGK}
\begin{eqnarray}
\left( D_{i}\partial _{j}V_{BH}\right) _{\frac{1}{2}-BPS} &=&\left( \partial
_{i}\partial _{j}V\partial _{BH}\right) _{\frac{1}{2}-BPS}=0;
\label{1/2-BPS-3} \\
\left( \partial _{i}\overline{\partial }_{\overline{j}}V_{BH}\right) _{\frac{%
1}{2}-BPS} &=&2g_{ij,\frac{1}{2}-BPS}\left| Z\right| _{\frac{1}{2}-BPS}^{2},
\label{1/2-BPS-4}
\end{eqnarray}
showing that there are no \textit{``flat'' directions} for such the ($\frac{1%
}{2}$-)BPS class of solutions to \textit{Attractor Eqs.} (\ref{N=2-d=4-AEs})
\cite{Ferrara-Marrani-2}.

\item  Non-supersymmetric (non-BPS) solutions to \textit{Attractor Eqs. }(%
\ref{N=2-d=4-AEs}) have $D_{i}Z\neq 0$ (\textit{at least}) for some $i\in
\left\{ 1,...,n-1\right\} $. Generally, such solutions fall into two class
\cite{BFGM1}, and they exhibit \textit{``flat'' directions} of $V_{BH}$
itself \cite{Ferrara-Marrani-2}. The non-BPS, $Z=0$ class is defined by the
following constraints:
\begin{equation}
D_{i}Z=\partial _{i}Z\neq 0,~\text{for~some~}i,~Z=0,~  \label{non-BPS-Z=0-1}
\end{equation}
thus yielding (from Eqs. (\ref{N=2-d=4-AEs}))
\begin{equation}
\left[ C_{ijk}g^{j\overline{j}}g^{k\overline{k}}\left( \overline{\partial }_{%
\overline{j}}\overline{Z}\right) \overline{\partial }_{\overline{k}}%
\overline{Z}\right] _{non-BPS,Z=0}=0.  \label{non-BPS-Z=0-2}
\end{equation}
Thus, Eqs. (\ref{VBH-2})\medskip\ and (\ref{non-BPS-Z=0-1}) yield
\begin{equation}
V_{BH,non-BPS,Z=0}=\left[ g^{i\overline{j}}\left( D_{i}Z\right) \overline{D}%
_{\overline{j}}\overline{Z}\right] _{non-BPS,Z=0}=\left[ g^{i\overline{j}%
}\left( \partial _{i}Z\right) \overline{\partial }_{\overline{j}}\overline{Z}%
\right] _{non-BPS,Z=0}.  \label{non-BPS-Z=0-3}
\end{equation}

\item  The non-BPS, $Z\neq 0$ class is defined by the following constraints:
\begin{equation}
D_{i}Z\neq 0,~\text{for~some~}i,~Z\neq 0. ~  \label{non-BPS-Z<>0-1}
\end{equation}
It is worth remarking that Eqs. (\ref{N=2-d=4-AEs}) and the non-BPS $Z\neq 0$
defining constraints (\ref{non-BPS-Z<>0-1}) imply the following relations to
hold at the non-BPS $Z\neq 0$ critical points of $V_{BH}$ \cite{ADFT-review}%
:
\begin{equation}
\left[ g^{i\overline{j}}\left( D_{i}Z\right) \overline{D}_{\overline{j}}%
\overline{Z}\right] _{non-BPS,Z\neq 0}=-\frac{i}{2}\left[ \frac{N_{3}\left(
\overline{Z}\right) }{\overline{Z}}\right] _{non-BPS,Z\neq 0}=\frac{i}{2}%
\left[ \frac{\overline{N}_{3}\left( Z\right) }{Z}\right] _{non-BPS,Z\neq 0},
\label{non-BPS-Z<>0-2}
\end{equation}
where the cubic form $N_{3}\left( \overline{Z}\right) $ is defined as \cite
{ADFT-review}
\begin{equation}
N_{3}\left( \overline{Z}\right) \equiv C_{ijk}\overline{Z}^{i}\overline{Z}%
^{j}\overline{Z}^{k}\Leftrightarrow \overline{N}_{3}\left( Z\right) \equiv
\overline{C}_{\overline{i}\overline{j}\overline{k}}Z^{\overline{i}}Z^{%
\overline{j}}Z^{\overline{k}}.  \label{N3}
\end{equation}
\medskip
\end{enumerate}

For an arbitrary SKG, it is in general hard to compute
\begin{equation}
\frac{S_{BH}}{\pi }=\left. V_{BH}\right| _{\partial _{\phi
}V_{BH}=0}=V_{BH}\left( \phi _{H}\left( \mathcal{P}\right) ,\mathcal{P}%
\right) ,
\end{equation}
where $\phi _{H}\left( \mathcal{P}\right) $ are the horizon scalar
configurations solving the \textit{Attractor Eqs.} (\ref{N=2-d=4-AEs}).
However, the situation dramatically simplifies for \textit{symmetric} SK
manifolds
\begin{equation}
\frac{G_{4}}{H_{4}},
\end{equation}
in which case a classification, analogous to the one available for $\mathcal{%
N}>2$-extended, $d=4$ supergravities (see \textit{e.g.} \cite{ADFT-review}
and Refs. therein; see also Sects. \ref{N=8-ungauged} and \ref{N=4-ungauged}%
) can be performed \cite{BFGM1}.

In the treatment below, we are going to give a remarkable general \textit{%
topological} formula for $V_{BH}\left( \phi _{H}\left( \mathcal{P}\right) ,%
\mathcal{P}\right) $ for \textit{symmetric} SKG, which is manifestly \textit{%
invariant} under diffeomorphisms of the SK scalar manifold, and which holds
\textit{for any} choice of symplectic basis of \textit{``bare''} charges $%
\mathcal{P}$ and of \textit{special coordinates} (see \textit{e.g.} \cite
{CDF-review} and Refs. therein) of the SK manifold itself. Indeed, such a
formula by no means does refer to \textit{special coordinates}, which may
not even exist for certain parametrizations of $\frac{G_{4}}{H_{4}}$ itself.

It should be pointed out that a general formula for the $G_{4}$-invariant $%
\mathcal{I}_{4,\mathcal{N}=2}$ is known for the so-called $d$-SK homogeneous
\textit{symmetric} manifolds \cite{dWVVP}, and it reads ($a=1,...,n-1$) \cite
{FG1}:
\begin{equation}
\mathcal{I}_{4,\mathcal{N}=2}\left( \mathcal{P}\right) =-\left(
p^{0}q_{0}+p^{a}q_{a}\right) ^{2}+4\left[ q_{0}\mathcal{I}_{3,\mathcal{N}%
=2}\left( p\right) -p^{0}\mathcal{I}_{3,\mathcal{N}=2}\left( q\right)
+\left\{ \mathcal{I}_{3,\mathcal{N}=2}\left( p\right) ,\mathcal{I}_{3,%
\mathcal{N}=2}\left( q\right) \right\} \right] ,  \label{I4-N=2-d=4-d=5}
\end{equation}
where
\begin{eqnarray}
\mathcal{I}_{3,\mathcal{N}=2}\left( p\right) &\equiv &\frac{1}{3!}%
d_{abc}p^{a}p^{b}p^{c};  \label{I4-N=2-d=4-d=5-1} \\
~\mathcal{I}_{3,\mathcal{N}=2}\left( q\right) &\equiv &\frac{1}{3!}%
d^{abc}q_{a}q_{b}q_{c}; \\
\left\{ \mathcal{I}_{3,\mathcal{N}=2}\left( p\right) ,\mathcal{I}_{3,%
\mathcal{N}=2}\left( q\right) \right\} &\equiv &\frac{\partial \mathcal{I}%
_{3,\mathcal{N}=2}\left( p\right) }{\partial p^{a}}\frac{\partial \mathcal{I}%
_{3,\mathcal{N}=2}\left( q\right) }{\partial q_{a}},
\label{I4-N=2-d=4-d=5-3}
\end{eqnarray}
in which the constant (number) rank-$3$ symmetric tensor $d_{abc}$ has been
introduced (and $d^{abc}$ is its suitably defined completely contravariant
form). However, such a formula holds for a particular symplectic basis
(namely the one inherited from the $\mathcal{N}=2$, $d=5$ theory, \textit{%
i.e.} the one of \textit{special coordinates}), in which the holomorphic
prepotential $F\left( X\right) $ of SKG can be written as
\begin{equation}
F\left( X\right) \equiv \frac{1}{3!}d_{abc}\frac{X^{a}X^{b}X^{c}}{X^{0}}.
\label{F-cubic}
\end{equation}
In such a symplectic basis, the manifest symmetry is the $d=5$ $U$-duality $%
G_{5}$, under which $G_{4}$ branches as $G_{4}\rightarrow G_{5}\times
SO\left( 1,1\right) $. Indeed, $\mathcal{I}_{3,\mathcal{N}=2}\left( p\right)
$ and $\mathcal{I}_{3,\mathcal{N}=2}\left( q\right) $ are nothing but
respectively the \textit{magnetic} and \textit{electric} invariants (both
\textit{cubic} in $\mathcal{P}$) of the relevant symplectic representations
of $G_{5}$.

Eq. (\ref{I4-N=2-d=4-d=5}) excludes the so-called \textit{quadratic} (or
\textit{minimally coupled} \cite{Luciani}) sequence of symmetric SK
manifolds (particular \textit{complex Grassmannians})
\begin{equation}
\frac{SU\left( 1,n-1\right) }{SU\left( n-1\right) \times U\left( 1\right) }%
,~n\in \mathbb{N}  \label{symm-quadr-seq}
\end{equation}
(not upliftable to $d=5$), for which $F\left( X\right) $ is given by (in the
symplectic basis exhibiting the maximal non-compact symmetry $SU\left(
1,n-1\right) $)
\begin{equation}
F\left( X\right) =-\frac{i}{2}\left[ \left( X^{0}\right)
^{2}-\sum_{i=1}^{n-1}\left( X^{i}\right) ^{2}\right] ,  \label{F-quadr-symm}
\end{equation}
and the invariant of the symplectic representation of $G_{4}=SU\left(
1,n-1\right) $ reads as follows (notice it is \textit{quadratic} in $%
\mathcal{P}$) \cite{ADF-U-duality-d=4}:
\begin{equation}
\mathcal{I}_{2,\mathcal{N}=2}\left( \mathcal{P}\right) =\left( p^{0}\right)
^{2}+q_{0}^{2}-\sum_{i=1}^{n-1}\left( \left( p^{i}\right)
^{2}+q_{i}^{2}\right) =\left| Z\right| ^{2}-g^{ij}\left( D_{i}Z\right)
\overline{D}_{\overline{j}}\overline{Z}.  \label{I2-symm-quadr-seq}
\end{equation}
Due to the \textit{quadratic} nature of the $G_{4}$-invariant $\mathcal{I}%
_{2,\mathcal{N}=2}\left( \mathcal{P}\right) $ given by Eq. (\ref
{I2-symm-quadr-seq}), the \textit{quadratic} sequence of symmetric SK
manifolds (\ref{symm-quadr-seq}) exhibits only one \textit{small} charge
orbit, namely the \textit{lightlike} one, beside the two \textit{large}
charge orbits determined in \cite{BFGM1}.

The symmetric SK manifolds whose geometry is determined by the holomorphic
prepotential function (\ref{F-cubic}) and the \textit{minimally coupled}
ones determined by Eq. (\ref{F-quadr-symm}) are \textit{all} the possible
symmetric SK manifolds. After \cite{CVP}, from the geometric perspective of
SKG, symmetric SK manifolds can be characterized in the following way.

In SKG the Riemann tensor obeys to the following constraint (see \textit{e.g.%
} \cite{CDF-review} and Refs. therein):
\begin{equation}
R_{i\overline{j}k\overline{l}}=-g_{i\overline{j}}g_{k\overline{l}}-g_{i%
\overline{l}}g_{k\overline{j}}+C_{ikm}\overline{C}_{\overline{l}\overline{j}%
\overline{n}}g^{m\overline{n}}.  \label{SKG-constraint}
\end{equation}
The requirement that the manifold to be symmetric demands the Riemann to be
covariantly constant:
\begin{equation}
D_{m}R_{i\overline{j}k\overline{l}}=0.  \label{Riemann-symm}
\end{equation}
Due to the SKG constraint (\ref{SKG-constraint}) and to covariant
holomorphicity of the $C$-tensor (expressed by Eq. (\ref{C-1})), Eq. (\ref
{Riemann-symm}) generally implies (for non-vanishing $C_{ijk}$)
\begin{equation}
D_{l}C_{ijk}=D_{(l}C_{i)jk}=0,  \label{C-symm}
\end{equation}
where in the last step Eq. (\ref{C-2}) was used. Thus, in a SK symmetric
space both the Riemann tensor and the $C$-tensor are covariantly constant.
Eq. (\ref{C-symm}) implies the following relation \cite{BFGM1}
\begin{equation}
C_{j(lm}C_{pq)k}\overline{C}_{\overline{i}\overline{j}\overline{k}}g^{j%
\overline{j}}g^{k\overline{k}}=\frac{4}{3}C_{(lmp}g_{q)\overline{i}},
\label{CCC=C}
\end{equation}
which is nothing but the \textit{``dressed''} form of the analogous relation
holding for the $d$-tensor itself \cite{GST,CVP}
\begin{equation}
d_{j(lm}d_{pq)k}d^{ijk}=\frac{4}{3}d_{(lmp}\delta _{q)}^{i}.  \label{d-symm}
\end{equation}
The quadratic sequence of symmetric manifolds (\ref{symm-quadr-seq}) whose
SKG is determined by the prepotential (\ref{F-quadr-symm}) has
\begin{equation}
C_{ijk}=0,  \label{C-symm-quadr-seq}
\end{equation}
whereas the remaning symmetric SK manifolds, whose prepotential in the
special coordinates is given by Eq. (\ref{F-cubic}) (with $d_{abc}$
constrained by Eq. (\ref{d-symm})), correspond to
\begin{equation}
C_{abc}=e^{K}d_{abc}.
\end{equation}

By using Eqs. (\ref{C-symm}) and (\ref{CCC=C}), as well as the SKG
identities (\ref{SKG-id-1})-(\ref{SKG-id-3}) (which, for symmetric SKG, are
equivalent to the \textit{Maurer-Cartan Eqs.}, as Eqs. (\ref{N=8-MC}) and (%
\ref{N=4-MC-1})-(\ref{N=4-MC-2}) for $\mathcal{N}=8$ and $\mathcal{N}=4$, $%
d=4$ supergravities, respectively; see \textit{e.g.} \cite
{ADF-U-duality-d=4,ADF-central}), one can prove that the following \textit{%
quartic} expression is a \textit{duality} invariant for all symmetric SK
manifolds :
\begin{eqnarray}
\mathcal{I}_{4,\mathcal{N}=2,symm}\left( \phi ,\mathcal{P}\right) &=&\left( Z%
\overline{Z}-Z_{i}\overline{Z}^{i}\right) ^{2}+  \notag \\
&&+\frac{2}{3}i\left( ZN_{3}\left( \overline{Z}\right) -\overline{Z}%
\overline{N}_{3}\left( Z\right) \right) +  \notag \\
&&-g^{i\overline{i}}C_{ijk}\overline{C}_{\overline{i}\overline{l}\overline{m}%
}\overline{Z}^{j}\overline{Z}^{k}Z^{\overline{l}}Z^{\overline{m}},
\label{I4-N=2-symm}
\end{eqnarray}
where the \textit{matter charges} have been re-noted as $Z_{i}\equiv D_{i}Z$%
, $Z^{\overline{i}}=g^{j\overline{i}}Z_{j}$, and definition (\ref{N3}) was
recalled.

As claimed above, $\mathcal{I}_{4,\mathcal{N}=2,symm}$ given by Eq. (\ref
{I4-N=2-symm}) is $\phi $-dependent \textit{only apparently}, \textit{i.e.}
it is \textit{topological}, merely charge-dependent:
\begin{equation}
\frac{\partial \mathcal{I}_{4,\mathcal{N}=2,symm}\left( \phi ,\mathcal{P}%
\right) }{\partial \phi }=0\Leftrightarrow \mathcal{I}_{4,\mathcal{N}%
=2,symm}=\mathcal{I}_{4,\mathcal{N}=2,symm}\left( \mathcal{P}\right) .
\end{equation}
Thus, by recalling Eq. (\ref{SBH-1}), the general entropy-area formula \cite
{Bek-Hawking} for extremal BHs in $\mathcal{N}=2$, $d=4$ supergravity
coupled to Abelian vector multiplets whose scalar manifold is a symmetric
(SK) space reads as follows:
\begin{equation}
\frac{S_{BH}}{\pi }=\left. V_{BH}\right| _{\partial _{\phi
}V_{BH}=0}=V_{BH}\left( \phi _{H}\left( \mathcal{P}\right) ,\mathcal{P}%
\right) =\left| \mathcal{I}_{4,\mathcal{N}=2,symm}\left( \mathcal{P}\right)
\right| ^{1/2}.  \label{SBH-N=2-1/2-BPS}
\end{equation}
\medskip

Let us briefly analyze Eq. (\ref{I4-N=2-symm}).

As for the case of $\mathcal{N}=8$, $d=4$ supergravity treated in Sect. \ref
{N=8-ungauged}, one can introduce a phase $\vartheta $ as follows (recall
definitions (\ref{N3})):
\begin{equation}
e^{2i\vartheta }\equiv -\frac{ZN_{3}\left( \overline{Z}\right) }{\overline{Z}%
\overline{N}_{3}\left( Z\right) }=\frac{iZC_{ijk}\overline{Z}^{i}\overline{Z}%
^{j}\overline{Z}^{k}}{-i\overline{C}_{\overline{l}\overline{m}\overline{n}%
}Z^{\overline{l}}Z^{\overline{m}}Z^{\overline{n}}}.
\end{equation}
Thus, $\vartheta $ is the phase of the quantity $iZN_{3}\left( \overline{Z}%
\right) $: $\vartheta \equiv \vartheta _{iZN_{3}\left( \overline{Z}\right) }$%
. It is then immediate to compute $\vartheta $ from Eq. (\ref{I4-N=2-symm}):
\begin{equation}
cos\vartheta \left( \phi ,\mathcal{P}\right) =\frac{3\left[ \mathcal{I}_{4,%
\mathcal{N}=2,symm}\left( \mathcal{P}\right) -\left( Z\overline{Z}-Z_{i}%
\overline{Z}^{i}\right) ^{2}+g^{i\overline{i}}C_{ijk}\overline{C}_{\overline{%
i}\overline{l}\overline{m}}\overline{Z}^{j}\overline{Z}^{k}Z^{\overline{l}%
}Z^{\overline{m}}\right] }{2^{2}\left| ZN_{3}\left( \overline{Z}\right)
\right| }.  \label{N=2-symm-phase}
\end{equation}
Notice that through Eq. (\ref{N=2-symm-phase}) ($cos$)$\vartheta $ is
determined in terms of the scalar fields $\phi $ and of the BH charges $%
\mathcal{P}$, also along the \textit{small} orbits where $\mathcal{I}_{4,%
\mathcal{N}=2,symm}=0$. However, Eq. (\ref{N=2-symm-phase}) is not defined
in the cases in which $ZN_{3}\left( \overline{Z}\right) =0$. In such cases, $%
\vartheta $ is actually undetermined. It should be clearly pointed out that
the phase $\vartheta $ has nothing to do with the phase of the $U\left(
1\right) $ bundle over the SK-Hodge vector multiplets' scalar manifold (see
\textit{e.g.} \cite{CDF-review} and Refs. therein).

\begin{enumerate}
\item  For $\frac{1}{2}$-BPS attractors (defined by the constraints (\ref
{1/2-BPS-1})), Eq. (\ref{I4-N=2-symm}) yields
\begin{equation}
\left. \mathcal{I}_{4,\mathcal{N}=2,symm}\right| _{\frac{1}{2}-BPS}=\left( Z%
\overline{Z}\right) _{\frac{1}{2}-BPS}^{2}=\left| Z\right| _{\frac{1}{2}%
-BPS}^{4},  \label{I4-N=2-1/2-BPS}
\end{equation}
as in turn also implied by Eqs. (\ref{1/2-BPS-2}) and (\ref{SBH-1}) (or
equivalently (\ref{SBH-N=2-1/2-BPS})). Notice that Eqs. (\ref{1/2-BPS-2})
and (\ref{I4-N=2-1/2-BPS}) are \textit{general}, \textit{i.e.} they hold for
any SKG, regardless the symmetric nature of the SK vector multiplets' scalar
manifold. Furthermore, the constraints (\ref{1/2-BPS-1}) imply that at the
event horizon of $\frac{1}{2}$-BPS extremal BHs it holds
\begin{equation}
\left[ N_{3}\left( \overline{Z}\right) \right] _{\frac{1}{2}%
-BPS}=0\Rightarrow \vartheta _{\frac{1}{2}-BPS}~\text{\textit{undetermined}}%
.~  \label{N=2-1/2-BPS-EH-phase}
\end{equation}

\item  For non-BPS $Z=0$ attractors (defined by the constraints (\ref
{non-BPS-Z=0-1}) which, through Eqs. (\ref{N=2-d=4-AEs}), imply Eq. (\ref
{non-BPS-Z=0-2})), Eq. (\ref{I4-N=2-symm}) yields
\begin{equation}
\left. \mathcal{I}_{4,\mathcal{N}=2,symm}\right| _{non-BPS,Z=0}=\left( Z_{i}%
\overline{Z}^{i}\right) _{non-BPS,Z=0}^{2}=\left[ g^{i\overline{j}}\left(
\partial _{i}Z\right) \overline{\partial }_{\overline{j}}\overline{Z}\right]
_{non-BPS,Z=0}^{2}.  \label{I4-N=2-non-BPS-Z=0}
\end{equation}
Notice that Eqs. (\ref{non-BPS-Z=0-3}) and (\ref{I4-N=2-non-BPS-Z=0}) are
\textit{general}, \textit{i.e.} they hold for any SKG, regardless the
symmetric nature of the SK vector multiplets' scalar manifold. Furthermore,
the constraints (\ref{1/2-BPS-1}) imply that at the event horizon of non-BPS
$Z=0$ extremal BHs it holds
\begin{equation}
Z_{non-BPS,Z=0}=0\Rightarrow \vartheta _{non-BPS,Z=0}~\text{\textit{%
undetermined}}.~  \label{N=2-non-BPS-Z=0-EH-phase}
\end{equation}

\item  For non-BPS $Z\neq 0$ attractors (defined by the constraints (\ref
{non-BPS-Z<>0-1}) as well as by Eqs. (\ref{N=2-d=4-AEs})), Eqs. (\ref
{I4-N=2-symm}) and (\ref{non-BPS-Z<>0-2}) yield
\begin{equation}
\left. \mathcal{I}_{4,\mathcal{N}=2,symm}\right| _{non-BPS,Z\neq
0}=-16\left| Z\right| _{non-BPS,Z\neq 0}^{4},  \label{Rule-of-3-1}
\end{equation}
thus implying, through Eq. (\ref{VBH-2}) \cite{TT-1,FK-N=8,BFGM1,ADFT-review}
\begin{equation}
\left. Z_{i}\overline{Z}^{i}\right| _{non-BPS,Z\neq 0}=3\left| Z\right|
_{non-BPS,Z\neq 0}^{2}\Leftrightarrow V_{BH,non-BPS,Z\neq 0}=4\left|
Z\right| _{non-BPS,Z\neq 0}^{2}.  \label{Rule-of-3-2}
\end{equation}
By plugging Eqs. (\ref{N=2-d=4-AEs}), (\ref{non-BPS-Z<>0-1}), (\ref
{non-BPS-Z<>0-2}) and (\ref{Rule-of-3-1}) into Eq. (\ref{N=2-symm-phase}),
it follows that at the event horizon of non-BPS $Z\neq 0$ extremal BHs it
holds that
\begin{equation}
\vartheta _{non-BPS,Z\neq 0}=\pi +2k\pi ,~k\in \mathbb{Z}.
\label{N=2-symm-non-BPS-Z<>0-EH-phase}
\end{equation}
It should be remarked that, differently from the results (\ref{1/2-BPS-2})-(%
\ref{1/2-BPS-4}), (\ref{I4-N=2-1/2-BPS})-(\ref{N=2-1/2-BPS-EH-phase})
(holding for $\frac{1}{2}$-BPS attractors) and from the results (\ref
{non-BPS-Z=0-2})-(\ref{non-BPS-Z=0-3}), (\ref{I4-N=2-non-BPS-Z=0})-(\ref
{N=2-non-BPS-Z=0-EH-phase}) (holding for non-BPS $Z=0$ attractors), Eqs. (%
\ref{Rule-of-3-1})-(\ref{N=2-symm-non-BPS-Z<>0-EH-phase}) are not \textit{%
general}: \textit{i.e.} they hold at the event horizon of extremal non-BPS $%
Z\neq 0$ BHs for symmetric SK manifolds, but they do not hold true for
generic SKG. However, when going \textit{beyond} the symmetric SK case (and
thus encompassing both homogeneous non-symmetric \cite
{dWVVP,DFT-hom-non-symm} and non-homogeneous SK spaces), one can compute
both $V_{BH,non-BPS,Z\neq 0}$ and $\left. \mathcal{I}_{4,\mathcal{N}%
=2,symm}\right| _{non-BPS,Z\neq 0}$, and express the deviation from the
symmetric case considered above in terms of the complex quantity \cite
{ADFT-review}
\begin{equation}
\Delta \equiv -\frac{3}{4}\frac{E_{i\overline{j}\overline{k}\overline{l}%
\overline{m}}\overline{Z}^{i}Z^{\overline{j}}Z^{\overline{k}}Z^{\overline{l}%
}Z^{\overline{m}}}{\overline{N}_{3}\left( Z\right) },
\end{equation}
where the tensor $E_{i\overline{j}\overline{k}\overline{l}\overline{m}}$ was
firstly introduced in \cite{dWVVP} (see also \cite{ADFT-review}). The
results of straightforward computations read as follows:
\begin{eqnarray}
V_{BH,non-BPS,Z\neq 0} &=&4\left| Z\right| _{non-BPS,Z\neq 0}^{2}+\Delta
_{non-BPS,Z\neq 0};  \label{VBH-N=2-non-symm-non-BPS-Z<>0} \\
\left. \mathcal{I}_{4,\mathcal{N}=2,symm}\right| _{non-BPS,Z\neq 0} &=&\left[
-16\left| Z\right| ^{4}+\Delta ^{2}-\frac{8}{3}\Delta \left| Z\right| ^{2}%
\right] _{non-BPS,Z\neq 0}.  \label{I4-N=2-non-symm-non-BPS-Z<>0}
\end{eqnarray}
Notice that, as yielded \textit{e.g.} by Eq. (\ref
{VBH-N=2-non-symm-non-BPS-Z<>0}), $\Delta $ is real at the non-BPS $Z\neq 0$
critical points of $V_{BH}$. For symmetric SK manifolds $E_{i\overline{j}%
\overline{k}\overline{l}\overline{m}}=0$ globally, and thus Eqs.(\ref
{VBH-N=2-non-symm-non-BPS-Z<>0}) and (\ref{I4-N=2-non-symm-non-BPS-Z<>0})
respectively reduce to Eqs. (\ref{Rule-of-3-2}) and (\ref{Rule-of-3-1}). On
the other hand, the results (\ref{Rule-of-3-1})-(\ref{Rule-of-3-2}) hold
also for those non-symmetry SK spaces ($E_{i\overline{j}\overline{k}%
\overline{l}\overline{m}}\neq 0$) such that
\begin{equation}
\Delta _{non-BPS,Z\neq 0}=0\Leftrightarrow \left( E_{i\overline{j}\overline{k%
}\overline{l}\overline{m}}\overline{Z}^{i}Z^{\overline{j}}Z^{\overline{k}}Z^{%
\overline{l}}Z^{\overline{m}}\right) _{non-BPS,Z\neq 0},
\label{non-symm-as-symm}
\end{equation}
where in the implication ``$\Rightarrow $'' the assumption $\left[ \overline{%
N}_{3}\left( Z\right) \right] _{non-BPS,Z\neq 0}\neq 0$ was made. The
condition (\ref{non-symm-as-symm}) might explain some results obtained for
generic ($d-$)SKGs in some particular supporting BH charge configurations in
\cite{TT-1} (see also the treatment in \cite{ADFT-review} and \cite
{Kallosh-review}).\medskip
\end{enumerate}

Consistently, for the quadratic \textit{minimally coupled} sequence (\ref
{symm-quadr-seq}), for which Eq. (\ref{C-symm-quadr-seq}) holds, Eq. (\ref
{I4-N=2-symm}) formally reduces to
\begin{gather}
\left. \mathcal{I}_{4,\mathcal{N}=2,symm}\right| _{C_{ijk}=0}=\left( Z%
\overline{Z}-Z_{i}\overline{Z}^{i}\right) ^{2};  \notag \\
\Updownarrow  \notag \\
\left| \left. \mathcal{I}_{4,\mathcal{N}=2,symm}\right| _{C_{ijk}=0}\right|
^{1/2}=\left| \mathcal{I}_{2,\mathcal{N}=2}\right| ,
\end{gather}
where $\mathcal{I}_{2,\mathcal{N}=2}$ is given by Eq. (\ref
{I2-symm-quadr-seq}).\medskip

Remarkably, Eq. (\ref{I4-N=2-symm}) turns out to be directly related to the
quantity $-h$ given by Eq. (2.31) of \cite{dWVVP} (see also the treatment of
\cite{Cecotti}). This is seen by noticing that Eq. (4.42) of \cite{dWVVP}
coincides with Eq. (\ref{I4-N=2-d=4-d=5}) (along with definitions (\ref
{I4-N=2-d=4-d=5-1})-(\ref{I4-N=2-d=4-d=5-3})). Note that the mapping of
quaternionic coordinates $\left( A^{\Lambda },B_{\Lambda }\right) ^{T}$ into
the charges $\mathcal{P}^{T}=\left( p^{\Lambda },q_{\Lambda }\right) ^{T}$
(in \textit{special coordinates}) is related to the $d=3$ attractor flows
(see \textit{e.g.} \cite{Quantum-AM,Gaiotto-Li-Padi,Trigiante-d=3-AM}%
).\medskip

For \textit{symmetric} SK manifolds, \textit{small }charge orbits of the
symplectic representation of $G_{4}$ are known to exist since \cite{FG1} and
\cite{LPS}.

\begin{itemize}
\item  \textit{small lightlike }charge orbits are defined by the $G_{4}$%
-invariant constraint
\begin{gather}
\mathcal{I}_{4,\mathcal{N}=2,symm}=0;  \label{N=2-symm-small-lightlike-1} \\
\Updownarrow  \notag \\
\left( Z\overline{Z}-Z_{i}\overline{Z}^{i}\right) ^{2}+\frac{2}{3}i\left(
ZN_{3}\left( \overline{Z}\right) -\overline{Z}\overline{N}_{3}\left(
Z\right) \right) =g^{i\overline{i}}C_{ijk}\overline{C}_{\overline{i}%
\overline{l}\overline{m}}\overline{Z}^{j}\overline{Z}^{k}Z^{\overline{l}}Z^{%
\overline{m}}.  \label{N=2-symm-small-lightlike-2}
\end{gather}
In this case, Eq. (\ref{N=2-symm-phase}) reduces to
\begin{equation}
\left. cos\vartheta \left( \phi ,\mathcal{P}\right) \right| _{\mathcal{I}_{4,%
\mathcal{N}=2,symm}=0}=\left. -\frac{3\left[ \left( Z\overline{Z}-Z_{i}%
\overline{Z}^{i}\right) ^{2}-g^{i\overline{i}}C_{ijk}\overline{C}_{\overline{%
i}\overline{l}\overline{m}}\overline{Z}^{j}\overline{Z}^{k}Z^{\overline{l}%
}Z^{\overline{m}}\right] }{2^{2}\left| ZN_{3}\left( \overline{Z}\right)
\right| }\right| _{\mathcal{I}_{4,\mathcal{N}=2,symm}=0}.
\end{equation}

\item  Beside the constraint (\ref{N=2-symm-small-lightlike-1})-(\ref
{N=2-symm-small-lightlike-2}), \textit{small critical }charge orbits are
defined by the following $G_{4}$-invariant set of first order differential
constraints, as well:
\begin{equation}
\frac{\partial \mathcal{I}_{4,\mathcal{N}=2,symm}}{\partial Z}=0=\frac{%
\partial \mathcal{I}_{4,\mathcal{N}=2,symm}}{\partial Z_{i}}.
\label{N=2-symm-small-critical-1}
\end{equation}

\item  Beside the constraints (\ref{N=2-symm-small-lightlike-1})-(\ref
{N=2-symm-small-lightlike-2}) and (\ref{N=2-symm-small-critical-1}), \textit{%
small doubly-critical }charge orbits are also defined by the following
set of second order differential constraints, as well:
\begin{equation}
\mathcal{D}_{i\overline{j}}\mathcal{I}_{4,\mathcal{N}=2,symm}=0=\mathcal{D}%
_{i}\mathcal{I}_{4,\mathcal{N}=2,symm},
\label{N=2-symm-small-doubly-critical-1}
\end{equation}
where the second-order differential operators $\mathcal{D}_{i\overline{j}}$
and $\mathcal{D}_{i}$ have been introduced:
\begin{eqnarray}
\mathcal{D}_{i\overline{j}} &\equiv &R_{i\overline{j}k}^{~~~l}\frac{\partial
}{\partial Z_{k}}\frac{\partial }{\overline{\partial }\overline{Z}^{l}};
\label{D_ijbar} \\
\mathcal{D}_{i} &\equiv &C_{ijk}\frac{\partial }{\partial Z_{j}}\frac{%
\partial }{\partial Z_{k}}.  \label{D_i}
\end{eqnarray}
Notice that, through the definitions (\ref{D_ijbar}) and (\ref{D_i}), the
constraints (\ref{N=2-symm-small-doubly-critical-1}) are $G_{4}$-invariant,
because they are equivalent to the following constraint:
\begin{equation}
\left. \frac{\partial ^{2}\mathcal{I}_{4,\mathcal{N}=2,symm}}{\partial Z_{%
\mathbf{sympl}\left( G_{4}\right) }\partial Z_{\mathbf{sympl}\left(
G_{4}\right) }}\right| _{\mathbf{Adj}\left( G_{4}\right) }=0,
\label{N=2-symm-small-double-criticality}
\end{equation}
where
\begin{equation}
Z_{\mathbf{sympl}\left( G_{4}\right) }\equiv \left( Z,\overline{Z}_{%
\overline{i}},\overline{Z},Z_{i}\right) ^{T},
\end{equation}
and the change of charge basis between the manifestly $H_{4}$-covariant (in
\textit{``flat'' local} coordinates) basis $Z_{\mathbf{sympl}\left(
G_{4}\right) }$ and the manifestly $Sp\left( 2n,\mathbb{R}\right) $%
-covariant basis $\mathcal{P}$ (defined by Eq. (\ref{P}))\ is expressed by
the fundamental \textit{identities} of the SKG (see \textit{e.g.} \cite
{N=2-big,CDF-review} and Refs. therein). Indeed, by considering the \textit{%
Cartan decomposition} of the Lie algebra of $G_{4}$:
\begin{equation}
\frak{g}_{4}=\frak{h}_{4}+\frak{k}_{4},
\end{equation}
and switching to \textit{``flat'' local} coordinates in the scalar manifold
(here denoted by capital Latin indices), it holds that $\mathcal{D}_{I}$ (%
\textit{``flat''} version of the operator defined in Eq. (\ref{D_i})) is $%
\frak{k}_{4}$-valued. Furthermore, in symmetric manifolds $R_{I\overline{J}%
K}^{~~~~L}$ is a two-form (in the first two \textit{``flat'' local} indices)
which is Lie algebra-valued in $\frak{h}_{4}$, and thus $\mathcal{D}_{I%
\overline{J}}$ (\textit{``flat''} version of the operator defined in Eq. (%
\ref{D_ijbar})) turns out to be $\frak{h}_{4}$-valued. Notice that Eq. (\ref
{N=2-symm-small-double-criticality}), $G_{4}$-invariantly defining the
\textit{small} \textit{doubly-critical} charge orbit(s) of the $\mathcal{%
N}=2$, $d=4$ vector multiplets' symmetric SK scalar manifolds, is the
analogue of Eq. (\ref{N=8-doubly-critical-4}), which defines in an $%
E_{7\left( 7\right) }$-invariant way the \textit{small} \textit{%
doubly-critical} charge orbit of $\mathcal{N}=8$, $d=4$ \textit{pure}
supergravity. It should be also recalled that in $\mathcal{N}=4$, $d=4$
\textit{matter coupled} supergravity \textit{small} \textit{%
doubly-critical} (or \textit{higher-order-critical}) charge orbits
(independent from the \textit{small} \textit{critical} ones) are absent.
As treated in Sect. \ref{N=4-ungauged}, all \textit{small} \textit{%
critical} charge orbits of the $\mathcal{N}=4$ theory actually are \textit{%
doubly-critical}, and the analogues of Eqs. (\ref{N=8-doubly-critical-4})
and (\ref{N=2-symm-small-double-criticality}) are given, through Eq. (\ref
{N=4-decomp}) and definitions (\ref{T-tensor-antisymm-2}) and (\ref{T0}), by
the rich case study exhibited by Eqs. ((\ref{A})-(\ref{B}) and) (\ref{A-1})-(%
\ref{B-1}).
\end{itemize}

The classification of \textit{small} charge orbits of the relevant
symplectic representation of $G_{4}$ for $\mathcal{N}=2$, $d=4$ supergravity
coupled to Abelian vector multiplets whose scalar manifold $\frac{G_{4}}{%
H_{4}}$ is (SK) symmetric, performed in accordance to their \textit{``order
of criticality''} (\textit{lightlike}, \textit{critical}, \textit{%
doubly-critical}), will be given elsewhere.

\section{\label{ADM-Mass}ADM Mass for BPS Extremal Black Hole States}

For BPS BH states in $d=4$ ungauged\footnote{In the present paper
only ungauged supergravities are treated. It is here worth remarking
that the definition of the \textit{ADM mass} for (eventually
rotating) asymptotically non-flat black holes in \textit{gauged}
supergravities is a fairly subtle issue, addressed by various
studies in literature (see \textit{e.g.} \cite{Gauged-1,Gauged-2},
and Refs. therein). } supergravity theories, the \textit{ADM mass}
\cite {ADM} $M_{ADM}\left( \phi _{\infty },\mathcal{P}\right) $ is
defined as the largest (of the absolute values) of the
\textit{skew-eigenvalues} of the (spatially asymptotically)
\textit{central charge matrix} $Z_{AB}\left( \phi _{\infty
},\mathcal{P}\right) $ which saturate the \textit{BPS bound} (\ref
{BPS-N}). The \textit{skew-diagonalization} of $Z_{AB}$ is made by
performing a suitable transformation of the
$\mathcal{R}$\textit{-symmetry}, and thus by going to the so-called
\textit{normal frame}. In such a frame, the
\textit{skew-eigenvalues} of $Z_{AB}$ can be taken to be real and
positive (up to an eventual overall \textit{phase}). By saturating
the \textit{BPS bound} (\ref{BPS-N}), it therefore holds that
\begin{equation}
M_{ADM}\left( \phi _{\infty },\mathcal{P}\right) =\left| \mathbf{Z}%
_{1}\left( \phi _{\infty },\mathcal{P}\right) \right| \geqslant ...\geqslant
\left| \mathbf{Z}_{\left[ \mathcal{N}/2\right] }\left( \phi _{\infty },%
\mathcal{P}\right) \right| ,
\end{equation}
where $\mathbf{Z}_{1}\left( \phi ,\mathcal{P}\right) ,...,\mathbf{Z}_{\left[
\mathcal{N}/2\right] }\left( \phi ,\mathcal{P}\right) $ denote the set of
\textit{skew-eigenvalues} of $Z_{AB}\left( \phi ,\mathcal{P}\right) $, and
square brackets denote the integer part of the enclosed number. As mentioned
at the end of Sect. \ref{Duality}, if $1\leqslant \mathbf{k}\leqslant \left[
\mathcal{N}/2\right] $ of the bounds expressed by Eq. (\ref{BPS-N}) are
saturated, the corresponding extremal BH state is named to be $\frac{\mathbf{%
k}}{\mathcal{N}}$-BPS. Thus, the minimal fraction of total supersymmetries
(pertaning to the asymptotically flat space-time metric) preserved by the
extremal BH background within the considered assumptions is $\frac{1}{%
\mathcal{N}}$ (for $\mathbf{k}=1$), while the maximal one is $\frac{1}{2}$
(for $\mathbf{k}=\frac{\mathcal{N}}{2}$).

The ADM mass and its symmetries are different, depending on $\mathbf{k}$.

\subsection{\label{N=8-ADM-Mass}$\mathcal{N}=8$}

In $\mathcal{N}=8$, $d=4$ supergravity (treated in Sect. \ref{N=8-ungauged}%
), the $E_{7\left( 7\right) }$ $U$-duality symmetry only allows the cases
\cite{FM} $\mathbf{k}=1,2,4$. By recalling the review given in Sect. \ref
{N=8-ungauged}, the maximal compact symmetries of the supporting charge
orbits respectively read \cite
{FM,FG1,FK-N=8,ADFT-review,Ferrara-Marrani-1,Ferrara-Marrani-2}
\begin{eqnarray}
\mathbf{k} &=&1:SU\left( 2\right) \times SU\left( 6\right) ; \\
\mathbf{k} &=&2:USp\left( 4\right) \times SU\left( 4\right) ; \\
\mathbf{k} &=&4:USp\left( 8\right) ,
\end{eqnarray}
and they hold all along the respective scalar flows. While cases $\mathbf{k}%
=2$ and $4$ are \textit{small }(thus not enjoying the \textit{attractor
mechanism}), case $\mathbf{k}=1$ can be either \textit{large} or \textit{%
small}.

In the \textit{large }$\mathbf{k}=1$ case, the \textit{attractor
mechanism} makes the maximal compact symmetry $SU\left( 2\right) \times
SU\left( 6\right) $ of the supporting charge orbit $\mathcal{O}_{\frac{1}{8}%
-BPS,\text{\textit{large}}}$ fully manifest as a symmetry of the central
charge matrix $Z_{AB}$ through the \textit{symmetry enhancement} (\ref
{1/8-BPS-large-enhancement}) at the event horizon of the considered extremal
BH.

Furthermore, the $\frac{1}{4}$-BPS saturation of the $\mathcal{N}=8$ BPS
bound (all along the $\frac{1}{4}$-BPS scalar flow) has the following
peculiar structure (recall Eq. (\ref{N=8-1/4-BPS-structure})) \cite{FM}
\begin{equation}
\left| \mathbf{Z}_{1}\left( \phi ,\mathcal{P}\right) \right| =\left| \mathbf{%
Z}_{2}\left( \phi ,\mathcal{P}\right) \right| >\left| \mathbf{Z}_{3}\left(
\phi ,\mathcal{P}\right) \right| =\left| \mathbf{Z}_{4}\left( \phi ,\mathcal{%
P}\right) \right| ,
\end{equation}
where it should be recalled that in Sect. \ref{N=8-ungauged} the notation $%
e_{i}\equiv \left| \mathbf{Z}_{i}\right| $ ($i=1,...,4$) was used.

As done in Sect. \ref{N=8-ungauged}, let us denote with $\lambda _{i}$ ($%
i=1,...,4$) the four real non-negative eigenvalues of the $8\times 8$
Hermitian matrix $Z_{AB}\overline{Z}^{CB}=\left( ZZ^{\dag }\right)
_{A}^{C}\equiv A_{A}^{C}$. Their relation with the absolute values of the
complex \textit{skew-eigenvalues} $e_{i}$ of $Z_{AB}$ is given by Eq. (\ref
{lambda-e}). As mentioned, the ordering $\lambda _{1}\geqslant \lambda
_{2}\geqslant \lambda _{3}\geqslant \lambda _{4}$ does not imply any loss of
generality.\smallskip\ After \cite{DFL-0-brane} (see in particular Eqs.
(4.74), (4.75), (4.86) and (4.87) therein), the explicit expression of $%
\lambda _{i}$ in terms of $U\left( 8\right) $-invariants (namely of $TrA$, $%
Tr\left( A^{2}\right) $, $Tr\left( A^{3}\right) $ and $Tr\left( A^{4}\right)
$, and suitable powers) is known, and it can be thus be used in order to
compute the ADM\ mass of $\frac{\mathbf{k}}{8}$-BPS extremal\textbf{\ }BH
states of $\mathcal{N}=8$, $d=4$ supergravity.

The $\lambda _{i}$'s are solution of the (square root of) \textit{%
characteristic equation} \cite{DFL-0-brane}
\begin{equation}
\sqrt{det\left( A-\lambda \mathbb{I}\right) }=\prod_{i=1}^{4}\left( \lambda
-\lambda _{i}\right) =\lambda ^{4}+a\lambda ^{3}+b\lambda ^{2}+c\lambda +d=0,
\label{N=8-d=4-char-eq}
\end{equation}
where \cite{DFL-0-brane}
\begin{eqnarray}
a &\equiv &-\frac{1}{2}TrA=-\left( \lambda _{1}+\lambda _{2}+\lambda
_{3}+\lambda _{4}\right) ;  \label{a} \\
&&  \notag \\
b &\equiv &\frac{1}{4}\left[ \frac{1}{2}\left( TrA\right) ^{2}-Tr\left(
A^{2}\right) \right] =  \notag \\
&=&\lambda _{1}\lambda _{2}+\lambda _{1}\lambda _{3}+\lambda _{1}\lambda
_{4}+\lambda _{2}\lambda _{3}+\lambda _{2}\lambda _{4}+\lambda _{3}\lambda
_{4};  \label{b} \\
&&  \notag \\
c &\equiv &-\frac{1}{6}\left[ \frac{1}{8}\left( TrA\right) ^{3}+Tr\left(
A^{3}\right) -\frac{3}{4}Tr\left( A^{2}\right) TrA\right] =  \notag \\
&=&-\left( \lambda _{1}\lambda _{2}\lambda _{3}+\lambda _{1}\lambda
_{2}\lambda _{4}+\lambda _{1}\lambda _{3}\lambda _{4}+\lambda _{2}\lambda
_{3}\lambda _{4}\right) ;  \label{c} \\
&&  \notag \\
d &\equiv &\frac{1}{4}\left[
\begin{array}{l}
\frac{1}{96}\left( TrA\right) ^{4}+\frac{1}{8}Tr^{2}\left( A^{2}\right) +%
\frac{1}{3}Tr\left( A^{3}\right) TrA+ \\
\\
-\frac{1}{2}Tr\left( A^{4}\right) -\frac{1}{8}Tr\left( A^{2}\right) Tr^{2}A
\end{array}
\right] =  \notag \\
&=&\sqrt{detA}=\lambda _{1}\lambda _{2}\lambda _{3}\lambda _{4}.  \notag \\
&&  \label{d}
\end{eqnarray}
The system (\ref{a})-(\ref{d}) can be inverted, yielding
\begin{eqnarray}
\lambda _{1,2} &=&-\frac{a}{4}+\frac{s}{2}\pm \frac{1}{2}\sqrt{\frac{a^{2}}{2%
}-\frac{4b}{3}-\frac{\left( a^{3}-4ab+8c\right) }{4s}-\frac{u}{3w}-\frac{w}{3%
}};  \label{lambda-1-2} \\
&&  \notag \\
\lambda _{3,4} &=&-\frac{a}{4}-\frac{s}{2}\pm \frac{1}{2}\sqrt{\frac{a^{2}}{2%
}-\frac{4b}{3}+\frac{\left( a^{3}-4ab+8c\right) }{4s}-\frac{u}{3w}-\frac{w}{3%
}},  \label{lambda-3-4}
\end{eqnarray}
where
\begin{eqnarray}
u &\equiv &b^{2}+12d-3ac;  \label{u} \\
v &\equiv &2b^{3}+27c^{2}-72bd-9abc+27a^{2}d;  \label{v} \\
w &\equiv &\left( \frac{v+\sqrt{v^{2}-4u^{3}}}{2}\right) ^{1/3};  \label{w}
\\
s &\equiv &\sqrt{\frac{a^{2}}{4}-\frac{2b}{3}+\frac{u}{3w}+\frac{w}{3}}.
\label{s}
\end{eqnarray}

Notice that the positivity of quantities under square root in Eqs. (\ref
{lambda-1-2}), (\ref{lambda-3-4}), (\ref{w}) and (\ref{s}) always holds.
Furthermore, Eq. (\ref{N=8-d=4-char-eq}) is at most of fourth order (for $%
\mathbf{k}=1$), of second order for $\mathbf{k}=2$, and of first order for $%
\mathbf{k}=1$.

\begin{enumerate}
\item  $\mathbf{k}=1$ ($\frac{1}{8}$-BPS, either \textit{large} or
\textit{small}). The $\frac{1}{8}$-BPS extremal BH square ADM mass is
\begin{equation}
M_{ADM,\frac{1}{8}-BPS}^{2}\left( \phi _{\infty },\mathcal{P}\right)
=\lambda _{1}\left( \phi _{\infty },\mathcal{P}\right) ,
\label{N=8-1/8-BPS-ADM-mass}
\end{equation}
where $\lambda _{1}$($>\lambda _{2}>\lambda _{3}>\lambda _{4}$, since $a<0$
and $s>0$) is given by Eq. (\ref{lambda-1-2}). In the \textit{large} $%
\mathbf{k}=1$ case $\lambda _{2}=$ $\lambda _{3}=\lambda _{4}=0$ at the
event horizon of the extremal BH, as given by Eq. (\ref{1/8-BPS-large-sol}).

\item  $\mathbf{k}=2$ ($\frac{1}{4}$-BPS, \textit{small}). As given by
Eq. (\ref{N=8-1/4-BPS-structure}), the eigenvalues are equal \textit{in pairs%
}. By suitably renaming the two non-coinciding $\lambda $'s, one gets
\begin{equation}
\lambda _{1,2}=\frac{1}{8}TrA\pm \frac{1}{2}\sqrt{\frac{1}{2}Tr\left(
A^{2}\right) -\frac{1}{16}\left( TrA\right) ^{2}}.  \label{lambdas}
\end{equation}
As mentioned above, the maximal (compact) symmetry is manifest when $\lambda
_{2}$ (in the renaming of Eq. (\ref{lambdas})) vanishes (see treatment in
Sect. \ref{N=8-ungauged}). Eq. (\ref{N=8-1/4-BPS-structure}) implies \cite
{DFL-0-brane}
\begin{eqnarray}
c &=&\frac{1}{2}a\left( b-\frac{1}{4}a^{2}\right) ;  \label{c-1/4-BPS} \\
d &=&\frac{1}{4}\left( b-\frac{1}{4}a^{2}\right) ^{2}.  \label{d-1/4-BPS}
\end{eqnarray}
In \cite{DFL-0-brane} Eqs. (\ref{c-1/4-BPS})-(\ref{d-1/4-BPS}) were shown to
be consequences of the criticality constraints (\ref{N=8-critical}). Thus,
the $\frac{1}{4}$-BPS extremal BH square ADM mass is
\begin{equation}
M_{ADM,\frac{1}{4}-BPS}^{2}\left( \phi _{\infty },\mathcal{P}\right)
=\lambda _{1}\left( \phi _{\infty },\mathcal{P}\right) ,
\label{N=8-1/4-BPS-ADM-mass-1}
\end{equation}
where $\lambda _{1}$($>\lambda _{2}$) is given by Eq. (\ref{lambdas}):
\begin{equation}
M_{ADM,\frac{1}{4}-BPS}^{2}\left( \phi _{\infty },\mathcal{P}\right) =\frac{1%
}{8}TrA\left( \phi _{\infty },\mathcal{P}\right) +\frac{1}{2}\sqrt{\frac{1}{2%
}Tr\left( A^{2}\right) \left( \phi _{\infty },\mathcal{P}\right) -\frac{1}{16%
}\left( TrA\left( \phi _{\infty },\mathcal{P}\right) \right) ^{2}}.
\label{N=8-1/4-BPS-ADM-mass-2}
\end{equation}

\item  $\mathbf{k}=4$ ($\frac{1}{2}$-BPS, \textit{small}). This case can
be obtained from the $\frac{1}{4}$-BPS considered at point 2 by further
putting $\lambda _{1}=\lambda _{2}$ in Eq. (\ref{lambdas}). Thus, \textit{all%
} eigenvalues of the Hermitian $8\times 8$ matrix $A$ are equal:
\begin{equation}
A_{A}^{C}=\frac{1}{8}\left( TrA\right) \delta _{A}^{C},
\end{equation}
which implies
\begin{equation}
Tr\left( A^{2}\right) =\frac{1}{8}\left( TrA\right) ^{2}.
\end{equation}
Therefore, $\frac{1}{2}$-BPS extremal BH square ADM mass is given by
\begin{equation}
M_{ADM,\frac{1}{2}-BPS}^{2}\left( \phi _{\infty },\mathcal{P}\right) =\frac{1%
}{8}TrA\left( \phi _{\infty },\mathcal{P}\right) =\frac{1}{16}Z_{AB}\left(
\phi _{\infty },\mathcal{P}\right) \overline{Z}^{AB}\left( \phi _{\infty },%
\mathcal{P}\right) .  \label{N=8-1/2-BPS-ADM-mass}
\end{equation}
\bigskip
\end{enumerate}

\subsection{\label{N=4-ADM-Mass}$\mathcal{N}=4$}

In $\mathcal{N}=4$, $d=4$ supergravity (treated in Sect. \ref{N=4-ungauged}%
), the $SL\left( 2,\mathbb{R}\right) \times SO\left( 6,M\right) $ $U$%
-duality symmetry only allows the cases \cite{FM} $\mathbf{k}=1,2$. By
recalling the treatment of Sect. \ref{N=4-ungauged}, the respective maximal
compact symmetries read \cite{FM,FG1,ADFT-review,Kallosh-review}
\begin{eqnarray}
\mathbf{k} &=&1:\left( SU\left( 2\right) \right) ^{2}\times SO\left(
M\right) \times SO\left( 2\right) ; \\
\mathbf{k} &=&2:USp\left( 4\right) \times SO\left( M\right) ,
\end{eqnarray}
and they hold all along the respective scalar flows. While case $\mathbf{k}%
=1 $ is \textit{large}, case $\mathbf{k}=2$ is \textit{small }(thus
not enjoying the \textit{attractor mechanism}).

In the \textit{large }$\mathbf{k}=1$ case, the \textit{attractor
mechanism} makes the maximal compact symmetry $\left( SU\left( 2\right)
\right) ^{2}\times SO\left( M\right) \times SO\left( 2\right) $ of the
supporting charge orbit $\mathcal{O}_{\frac{1}{4}-BPS,\text{large}}$ fully
manifest as a symmetry of the central charge matrix $Z_{AB}$ through the
\textit{symmetry enhancement} (recall Eq. (\ref
{N=4-1/4-BPS-large-enhancement}))
\begin{equation}
\left( SU\left( 2\right) \right) ^{2}\times SO\left( M-2\right) \times
SO\left( 2\right) \overset{r\rightarrow r_{H}^{+}}{\longrightarrow }\left(
SU\left( 2\right) \right) ^{2}\times SO\left( M\right) \times SO\left(
2\right)
\end{equation}
at the event horizon of the considered extremal BH.

As done in Sect. \ref{N=4-ungauged} and in the treatment of case $\mathcal{N}%
=8$, $d=4$ above, let us denote with $\lambda _{1}$ and $\lambda _{2}$ the
two real non-negative eigenvalues of the $4\times 4$ Hermitian matrix $Z_{AB}%
\overline{Z}^{CB}=\left( ZZ^{\dag }\right) _{A}^{C}\equiv A_{A}^{C}$. Their
relation with the absolute values of the complex \textit{skew-eigenvalues} $%
e_{i}$ of $Z_{AB}$ is given by Eq. (\ref{lambda-e}). As mentioned, the
ordering $\lambda _{1}\geqslant \lambda _{2}$ does not imply any loss of
generality.\smallskip\ After \cite{DFL-0-brane}, the explicit expression of $%
\lambda _{1}$ and $\lambda _{2}$ in terms of $\left( U\left( 4\right) \times
SO\left( M\right) \right) $-invariants (namely of $TrA$, $Tr\left(
A^{2}\right) $ and $\left( TrA\right) ^{2}$) is known, and it can be thus be
used in order to compute the ADM\ mass of $\frac{\mathbf{k}}{4}$-BPS extremal%
\textbf{\ }BH states of $\mathcal{N}=4$, $d=4$ supergravity.

Indeed, $\lambda _{1}$ and $\lambda _{2}$ are solution of the (square root
of) \textit{characteristic equation} \cite{DFL-0-brane}
\begin{equation}
\sqrt{det\left( A-\lambda \mathbb{I}\right) }=\prod_{i=1}^{2}\left( \lambda
-\lambda _{i}\right) =\lambda ^{2}-\frac{1}{2}\left( TrA\right) \lambda
+\left( detA\right) ^{1/2}=0,  \label{N=4-d=4-char-eq}
\end{equation}
whose solution reads
\begin{equation}
\lambda _{1,2}=\frac{1}{2}\left( \frac{1}{2}TrA\pm \sqrt{Tr\left(
A^{2}\right) -\frac{1}{4}\left( TrA\right) ^{2}}\right) .
\label{N=4-lambdas}
\end{equation}

Notice that the positivity of quantities under square root in Eq. (\ref
{N=4-lambdas}) always holds. Furthermore, Eq. (\ref{N=4-d=4-char-eq}) is at
most of second order (for $\mathbf{k}=1$) and of first order for $\mathbf{k}%
=2$.

\begin{enumerate}
\item  $\mathbf{k}=1$ ($\frac{1}{4}$-BPS \textit{large}). The $\frac{1}{4%
}$-BPS extremal BH square ADM mass is
\begin{eqnarray}
M_{ADM,\frac{1}{4}-BPS}^{2}\left( \phi _{\infty },\mathcal{P}\right)
&=&\lambda _{1}\left( \phi _{\infty },\mathcal{P}\right) =  \notag \\
&=&\frac{1}{2}\left( \frac{1}{2}TrA\left( \phi _{\infty },\mathcal{P}\right)
+\sqrt{Tr\left( A^{2}\right) \left( \phi _{\infty },\mathcal{P}\right) -%
\frac{1}{4}\left( TrA\left( \phi _{\infty },\mathcal{P}\right) \right) ^{2}}%
\right) ,  \notag \\
&&  \label{N=4-d=4-1/4-BPS-ADM-Mass}
\end{eqnarray}
where $\lambda _{1}>\lambda _{2}$. Notice that $\lambda _{2}=0$ at the event
horizon of the extremal BH, as given by Eq. (\ref{N=4-1/4-BPS-large-sol-1}).

\item  $\mathbf{k}=2$ ($\frac{1}{2}$-BPS, \textit{small}). This case can
be obtained from the $\frac{1}{4}$-BPS considered at point 1 by further
putting $\lambda _{1}=\lambda _{2}$ in Eq. (\ref{N=4-lambdas}). Thus,
\textit{all} eigenvalues of the Hermitian $4\times 4$ matrix $A$ are equal:
\begin{equation}
A_{A}^{C}=\frac{1}{4}\left( TrA\right) \delta _{A}^{C},
\end{equation}
which implies
\begin{equation}
Tr\left( A^{2}\right) =\frac{1}{4}\left( TrA\right) ^{2}.
\end{equation}
Thus, the $\frac{1}{2}$-BPS extremal BH square ADM mass is
\begin{equation}
M_{ADM,\frac{1}{2}-BPS}^{2}\left( \phi _{\infty },\mathcal{P}\right)
=\lambda _{1}\left( \phi _{\infty },\mathcal{P}\right) =\lambda _{2}\left(
\phi _{\infty },\mathcal{P}\right) =\frac{1}{4}TrA\left( \phi _{\infty },%
\mathcal{P}\right) .
\end{equation}
\bigskip
\end{enumerate}

It should be here remarked that the $\mathcal{R}$\textit{-symmetry} of the $%
\frac{\mathbf{k}}{\mathcal{N}}$-BPS extremal BH states, \textit{i.e.} the
compact symmetry of the solution in the \textit{normal frame} (determining
the automorphism group of the supersymmetry algebra in the \textit{rest frame%
}) gets broken as follows:
\begin{equation}
\mathcal{R}\longrightarrow USp\left( 2\mathbf{k}\right) \times ...~.
\end{equation}
This is precisely the symmetry of the
$\frac{\mathbf{k}}{\mathcal{N}}$-BPS saturated massive multiplets of
the $\mathcal{N}$-extended, $d=4$ Poincar\'{e} supersymmetry algebra
\cite{FSZ}.\bigskip

We end this Section by finally commenting about the ADM mass for non-BPS
extremal BH states.

In non-BPS cases, ADM mass of extremal BH states is not directly related to
the \textit{skew-eigenvalues} of the \textit{central charge matrix} $Z_{AB}$%
. For some non-BPS extremal BHs a \textit{``fake supergravity (first order)
formalism''} \cite{Fake-Refs} can be consistently formulated in terms of a
\textit{``fake superpotential''} $\mathcal{W}\left( \phi ,\mathcal{P}\right)
$ \cite{CD,ADOT-1,Gnecchi-1,Perz-first-order} such that (also recall Eq. (%
\ref{SBH-1}))
\begin{eqnarray}
\left. \mathcal{W}_{non-BPS}^{2}\left( \phi ,\mathcal{P}\right) \right| _{%
\frac{\partial \mathcal{W}}{\partial \phi }=0} &\equiv &\mathcal{W}%
_{non-BPS}^{2}\left( \phi _{H,non-BPS}\left( \mathcal{P}\right) ,\mathcal{P}%
\right) =  \notag \\
&=&\left. V_{BH}\left( \phi ,\mathcal{P}\right) \right| _{\frac{\partial
V_{BH}}{\partial \phi }=0}\equiv V_{BH}\left( \phi _{H,non-BPS}\left(
\mathcal{P}\right) ,\mathcal{P}\right) =  \notag \\
&=&\frac{S_{BH,non-BPS}\left( \mathcal{P}\right) }{\pi },
\end{eqnarray}
with $\mathcal{W}_{non-BPS}$ varying, dependently on whether $Z_{AB}=0$ or
not. In such frameworks, the general expression of the non-BPS ADM mass
reads as follows \cite{CD,ADOT-1,Gnecchi-1}
\begin{equation}
M_{ADM,non-BPS}\left( \phi _{\infty },\mathcal{P}\right) =\mathcal{W}%
_{non-BPS}\left( \phi _{\infty },\mathcal{P}\right) .
\end{equation}

\section*{Acknowledgments}

This work is supported in part by the ERC Advanced Grant no. 226455, \textit{%
``Supersymmetry, Quantum Gravity and Gauge Fields''}
(\textit{SUPERFIELDS}).

We would like to thank M. Trigiante for enlightening discussions.

A. M. would like to thank the CTP of the University of California, Berkeley,
CA USA, the Department of Physics, University of Cincinnati, OH USA, and the
Department of Physics, Theory Unit Group at CERN, Geneva CH, where part of
this work was done, for kind hospitality and stimulating environment.

The work of B. L. C. and B. Z. ~has been supported in part by the Director,
Office of Science, Office of High Energy and Nuclear Physics, Division of
High Energy Physics of the U.S. Department of Energy under Contract No.
DE-AC02-05CH11231, and in part by NSF grant 10996-13607-44 PHHXM.

A substantial part of S. F.'s investigation was performed at the Center for
Theoretical Physics (CTP), University of California, Berkeley, CA USA, with
S. F. sponsored by a \textit{``Miller Visiting Professorship"} awarded by
the Miller Institute for Basic Research on Science. The work of S. F.~has
been supported also in part by INFN - Frascati National Laboratories, and by
D.O.E.~grant DE-FG03-91ER40662, Task C.

The work of A. M. has been supported by an INFN visiting Theoretical
Fellowship at SITP, Stanford University, Stanford, CA, USA.

\end{document}